\documentclass[aps,prl,reprint,superscriptaddress,longbibliography,floatfix]{revtex4-2}

\usepackage{amsmath}
\usepackage{amssymb}
\usepackage{amsfonts}
\usepackage{bm}
\usepackage{graphicx}
\usepackage{xcolor}
\usepackage{placeins}
\usepackage{tikz}
\usetikzlibrary{arrows.meta,positioning,decorations.pathmorphing,calc,patterns}

\makeatletter
\let\cat@comma@active\@empty
\makeatother

\begin{document}
\title{Reservoir-Engineered Josephson Diode in a Dark Andreev Channel}

\author{Jianxiong Zhai}
\affiliation{Shanghai Institute of Microsystem and Information Technology,
Chinese Academy of Sciences, Shanghai 200050, China}
\affiliation{University of Chinese Academy of Sciences, Beijing 100049, China}

\author{Zelei Zhang}
\affiliation{Shanghai Institute of Microsystem and Information Technology,
Chinese Academy of Sciences, Shanghai 200050, China}
\affiliation{School of Physical Science and Technology, ShanghaiTech University, Shanghai 201210, China}

\author{Jiawei Yan}
\email{yanjw@mail.sim.ac.cn}
\affiliation{Shanghai Institute of Microsystem and Information Technology,
Chinese Academy of Sciences, Shanghai 200050, China}
\affiliation{University of Chinese Academy of Sciences, Beijing 100049, China}

\date{\today}

\begin{abstract}
We propose a reservoir-engineered Josephson diode in a double-dot junction with symmetric coherent couplings.
A relative phase in the coupling to a common loss reservoir selects a rapidly decaying bright orbital and an orthogonal dark Andreev channel.
In the quantum Zeno regime, virtual bright-sector transitions set the dark-channel occupation while preserving its coherent supercurrent.
For junctions in the infinite-gap Zeno regime, the steady-state current factorizes as $I_{\rm ss}=I_A P_\gamma$: a reciprocal spectral current $I_A$ is reweighted by the reservoir-selected occupation imbalance $P_\gamma$, which generates rectification.
The resulting nonreciprocal current remains finite as reservoir leakage decreases with increasing loss strength.
Furthermore, local Coulomb repulsion tunes a dark-pair resonance that enables near-ideal rectification, with induced two-particle loss essential to the quantitative description of the steady current.
Keldysh-Lindblad calculations confirm that the results remain robust at finite loss strength and superconducting gap.
\end{abstract}

\maketitle

\paragraph{Introduction.---}
A Josephson diode is a superconducting weak link with unequal forward and backward critical currents, $I_c^+\neq |I_c^-|$, enabling nonreciprocal supercurrent transport~\cite{Fulton1974PRB,Nadeem2023NatRevPhys,Jiang2022NatPhys}.
In equilibrium, such nonreciprocity requires breaking both inversion and time-reversal symmetry~\cite{Zhang2022PRX}.
Established mechanisms include finite-momentum pairing and spin-dependent couplings~\cite{Yuan2022PNAS,Daido2022PRL,Davydova2022SciAdv,Ando2020Nature,IlicBergeret2022PRL,MeyerHouzet2024APL,IlicVirtanen2024PRB}, anomalous phase shifts combined with higher harmonics~\cite{Buzdin2008PRL,Reynoso2008PRL,Yokoyama2014PRB,Strambini2020NatNano,Baumgartner2022NatNano,LeggLossKlinovaja2022PRB}, and geometric or multiterminal interference~\cite{Lyu2021NatCommun,Golod2022NatCommun,Gupta2023NatCommun,SeoaneSouto2022PRL}.
In these settings, rectification is governed by the phase dependence of the junction free energy.
Out of equilibrium, Andreev-state occupations need no longer follow the Fermi-Dirac distribution and provide an additional means of controlling the supercurrent~\cite{Beenakker1991PRL,Morpurgo1998APL,Pala2007NJP}.
Nonequilibrium control has enabled supercurrent reversal in controllable $\pi$ junctions~\cite{Baselmans1999Nature} and nonreciprocal currents in periodically driven SNS junctions~\cite{Soori2023PhysScr}, symmetric Andreev interferometers carrying a dissipative control current~\cite{Shaffer2025PRB}, and quantum-dot triodes with unequal superconducting couplings~\cite{Li2026arXiv}.
These results motivate exploring reservoir control of Andreev-state occupations as a route to Josephson nonreciprocity.

Reservoir engineering controls relaxation pathways and steady-state occupations by designing system-environment couplings, turning dissipation into a resource for state preparation~\cite{Poyatos1996PRL,Kraus2008PRA,Diehl2008NatPhys,Verstraete2009NatPhys,Barreiro2011Nature,Schindler2013NatPhys,Diehl2011NatPhys,Bardyn2013NJP,Metelmann2015PRX}.
In a Josephson junction, however, coupling to a particle reservoir can suppress the supercurrent~\cite{Beenakker2024APL}. 
The challenge is therefore to use engineered loss to control Andreev-state occupations while preserving coherent pair transport.
The quantum Zeno effect provides a route around this obstacle by suppressing transitions into the lossy sector~\cite{Misra1977JMP,Facchi2009JPhysConf}.
Suppose particles can leave the system only after entering a lossy collective mode. 
With coherent coupling $g$ and decay rate $\kappa \gg g$, the mode decays much faster than it is populated and therefore remains nearly empty, with an occupation of order $(g/\kappa)^2$.
The resulting particle-loss rate is therefore $\kappa(g/\kappa)^2 = g^2/\kappa$.
Thus, increasing the damping suppresses, rather than enhances, the total particle loss.
Strong dissipation confines the slow dynamics mainly to the dark subspace orthogonal to the lossy mode~\cite{Facchi2002PRA,Facchi2002PRL,Zanardi2014PRL}, a counterintuitive effect also observed with ultracold molecules~\cite{Syassen2008Science}.

\begin{figure}
    \centering
    \includegraphics[width=\columnwidth]{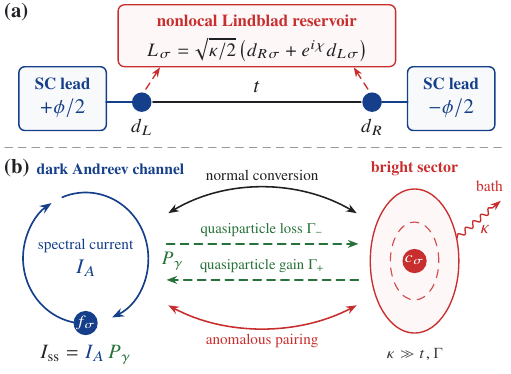}
    \caption{Zeno selection and occupation control of a dark Andreev channel.
    (a) Two dots ($d_L,d_R$) coupled by hopping $t$ and proximitized by superconducting leads with phases $\pm\phi/2$.
    A nonlocal reservoir with jump operator $L_\sigma=\sqrt{\kappa/2}(d_{R\sigma}+e^{i\chi}d_{L\sigma}):=\sqrt{\kappa}c_\sigma$ damps the bright orbital.
    (b) In the Zeno regime, $\kappa\gg t,\Gamma$, the bright orbital $c_\sigma$ is rapidly damped, while the dark orbital $f_\sigma$ supports the spectral current $I_A$.
    Double-headed arrows denote coherent normal conversion and anomalous pairing between bright and dark orbitals.
    After eliminating the rapidly decaying bright orbital, these couplings generate effective loss and gain of dark Andreev quasiparticles at rates $\Gamma_-$ and $\Gamma_+$ (green arrows).
    Their balance fixes $P_\gamma=1-2n_\gamma$ and hence the leading current $I_{\rm ss}=I_AP_\gamma$ [Eq.~\eqref{eq:intro_factorized}].}
    \label{fig:concept}
\end{figure}

In this work, we implement this idea in a double-dot Josephson junction with symmetric coherent couplings and a phase-textured common reservoir, as sketched in Fig.~\ref{fig:concept}.
The reservoir damps only the collective orbital $d_{R\sigma}+e^{i\chi}d_{L\sigma}$ (bright mode), with $\chi$ the relative coupling phase controlled by the magnetic flux through the loss loop.
In the quantum Zeno regime, where dissipation is the largest scale, the orthogonal dark mode survives as an Andreev channel whose supercurrent remains of the order of that of the lossless junction, and it does not vanish as the dissipation grows.
Integrating out the bright sector generates a competing quasiparticle loss and gain in the dark channel, and their balance fixes the occupation $n_\gamma$ of the dark Bogoliubov quasiparticle.
For the noninteracting junction in the infinite-gap Zeno limit, a secular reduction gives the leading steady current as
\begin{equation}
I_{\rm ss}(\phi)=I_A(\phi)P_\gamma(\phi),
\label{eq:intro_factorized}
\end{equation}
where $I_A$ is the spectral current of the dark Andreev channel and
$P_\gamma=1-2n_\gamma \in [-1,1]$ is its occupation imbalance.
Eq.~\eqref{eq:intro_factorized} is the central result of this work: $I_A$ is strictly reciprocal, meaning that its positive and negative critical currents have equal magnitude, so all diode nonreciprocity is written by the reservoir through occupation $P_\gamma$.
We further show that local Coulomb repulsion enhances rectification through a dark-pair resonance, where an induced two-particle loss channel is essential and the diode efficiency approaches unity with polarity set by the reservoir phase.
Finally, Keldysh-Lindblad calculations confirm the robustness of the above results at finite $\kappa$ and superconducting gap.

The rest of the Letter proceeds in three stages: we first solve the noninteracting problem analytically in the Zeno limit, then incorporate Coulomb interactions and benchmark against exact numerics, and finally go beyond the strict Zeno and infinite-gap limits using Keldysh-Lindblad Green's function theory.
Detailed derivations are provided in the Supplemental Material (SM), which is included below.

\paragraph{Model and theory.---}
We consider a double-dot system proximitized by superconducting leads, as shown in Fig.~\ref{fig:concept}~(a).
For low-energy physics inside the SC gap, the effective Hamiltonian reads \cite{Meng2009PRB,Beenakker2024APL}
$H=H_t+H_\Delta$, with
$H_t=t\sum_\sigma(d^\dagger_{L\sigma}d_{R\sigma}+{\rm H.c.})$ and
$H_\Delta=\Gamma_L e^{i\phi/2}d^\dagger_{L\uparrow}d^\dagger_{L\downarrow}+\Gamma_R e^{-i\phi/2}d^\dagger_{R\uparrow}d^\dagger_{R\downarrow}+{\rm H.c.}$~\cite{PhysRevB.103.235163,zhang2026manybodyjosephsondiodeeffect}.
Here, $t$ is the interdot tunneling amplitude, $\Gamma_{L,R}$ are the induced pairing amplitudes, and the superconducting phases are chosen symmetrically as $\phi_{L,R}=\pm\phi/2$.
We set $\Gamma := \Gamma_L = \Gamma_R$ unless explicitly stated otherwise.
The junction is coupled to an engineered reservoir, and the density matrix evolves according to the Lindblad master equation~\cite{Lindblad1976CMP,Gorini1976JMP}
\begin{equation}
\dot{\rho}
=
-i[H,\rho]
+
\sum_\sigma \mathcal{D}[L_\sigma]\rho,
\quad
\mathcal{D}[L]\rho
=
L\rho L^\dagger
-\frac{1}{2}\{L^\dagger L,\rho\},
\label{eq:lindblad}
\end{equation}
with the phase-textured nonlocal jump operator
\begin{equation}
L_\sigma
=
\sqrt{\frac{\kappa}{2}}
\left(
d_{R\sigma}
+
e^{i\chi}d_{L\sigma}
\right),
\label{eq:nonlocal_jump}
\end{equation}
where $\kappa$ is the loss rate and $\chi$ is the relative phase between the reservoir couplings to the left and right dots, controlled by the flux through the loss loop.
This nonlocal jump operator can arise microscopically when both dots tunnel into a fast-decaying auxiliary mode (SM, Sec.~\ref{sec:SM_auxiliary}).
Local gauge transformations can redistribute phases among $H_t$, $H_\Delta$, and $L_\sigma$, so only gauge-invariant combinations can enter physical observables,
and $\eta := \phi/2 + \chi$ is a gauge-invariant quantity that encodes these relative phases (SM, Sec.~\ref{sec:SM_bright_dark}).

\paragraph{Zeno-projected dark channel.---}
We introduce the bright-dark basis $c_\sigma=(d_{R\sigma}+e^{i\chi}d_{L\sigma})/\sqrt{2}$ and $f_\sigma=(-e^{-i\chi}d_{R\sigma}+d_{L\sigma})/\sqrt{2}$, for which Eq.~\eqref{eq:nonlocal_jump} becomes $L_\sigma=\sqrt{\kappa}c_\sigma$.
The reservoir directly damps the bright orbital $c_\sigma$, while $f_\sigma$ lies in the dark subspace of the bare dissipator.
When $\kappa$ is much larger than $t$ and $\Gamma_{L,R}$, the rapid decay of the bright orbital suppresses transitions into it through the quantum Zeno effect.
The slow dynamics is therefore confined predominantly to the dark orbital, which can retain phase coherence and carry supercurrent.

The original Hamiltonian can be rewritten in the bright-dark basis as
$H=H_c+H_f+H_{cf}$.
Here,
\begin{equation}
H_f =-t\cos\chi\sum_\sigma f^\dagger_\sigma f_\sigma+(\Gamma e^{-i\chi}\cos\eta f^\dagger_\uparrow f^\dagger_\downarrow+{\rm H.c.}),
\end{equation}
and
$
H_c =t\cos\chi\sum_\sigma c^\dagger_\sigma c_\sigma+(\Gamma e^{i\chi}\cos\eta c^\dagger_\uparrow c^\dagger_\downarrow+{\rm H.c.})
$
are the dark and bright sector Hamiltonians, respectively, and
$
H_{cf} = -i t e^{i\chi}\sin\chi
(c^\dagger_\uparrow f_\uparrow+c^\dagger_\downarrow f_\downarrow)
+
i\Gamma\sin\eta(c^\dagger_\uparrow f^\dagger_\downarrow-c^\dagger_\downarrow f^\dagger_\uparrow)
+
{\rm H.c.}
$
is the hybridization between them.
In the Zeno regime $\kappa\gg t,\Gamma$, $H_f$ describes the surviving Andreev transport channel, whereas $H_{cf}$ couples it to the short-lived bright sector through normal conversion proportional to $t\sin\chi$ and anomalous bright-dark pair creation proportional to $\Gamma\sin\eta$.
The bare low-frequency propagator of a bright fermion is then
$\mathsf g_{c,\sigma\sigma'}^R(0)=-2i\delta_{\sigma\sigma'}/\kappa+O(\kappa^{-2})$.
Connecting two $H_{cf}$ vertices through this propagator, or equivalently eliminating the bright sector to second order~\cite{Reiter2012PRA}, gives the effective dark-sector dynamics
\begin{equation}
\dot\rho_f=-i[H_f,\rho_f]+\sum_{\mu=1,2}\mathcal D[L_{f,\mu}]\rho_f,
\label{eq:dark_lindblad_main}
\end{equation}
with
$
L_{f,1/2}=\sqrt{{4}/{\kappa}}\,(t\sin\chi f_{\uparrow/\downarrow}\mp\Gamma e^{-i\chi}\sin\eta f^\dagger_{\downarrow/\uparrow})
$.
Although $f_\sigma$ is dark to the bare jump operator in Eq.~\eqref{eq:nonlocal_jump}, a virtual bright-sector process combines a propagator proportional to $1/\kappa$ with the bright loss amplitude proportional to $\sqrt{\kappa}$.
The resulting jump amplitudes $L_{f,\mu}\propto\kappa^{-1/2}$ therefore produce dissipative terms of order $1/\kappa$.

To identify how this projected dissipation sets the Andreev-level occupation, we diagonalize $H_f$ using
$f_\uparrow=u\gamma_\uparrow+e^{-i\chi}v\gamma^\dagger_\downarrow$ and
$f_\downarrow=u\gamma_\downarrow-e^{-i\chi}v\gamma^\dagger_\uparrow$.
The dark Andreev energy is
$E_A=\sqrt{t^2\cos^2\chi+\Gamma^2\cos^2\eta}$,
and the real coherence factors satisfy
$u^2+v^2=1$,
$u^2-v^2=t\cos\chi/E_A$,
and
$2uv=\Gamma\cos\eta/E_A$.
In this Bogoliubov basis, the same projected jump operators become
\begin{equation}
L_{f,1/2}=\sqrt{\frac{4}{\kappa}}\left(A_-\gamma_{\uparrow/\downarrow}\mp e^{-i\chi}A_+\gamma^\dagger_{\downarrow/\uparrow}\right),
\label{eq:rates_main}
\end{equation}
where $A_-=t\sin\chi u+\Gamma\sin\eta v$ and $A_+=\Gamma\sin\eta u-t\sin\chi v$ are the quasiparticle annihilation and creation amplitudes, respectively.

Substituting Eq.~\eqref{eq:rates_main} into Eq.~\eqref{eq:dark_lindblad_main} separates the projected dissipator into quasiparticle loss terms with rate $\Gamma_- = 4A_-^2/\kappa$, gain terms with rate $\Gamma_+ = 4A_+^2/\kappa$, and interference terms that generate pair coherences.
The interference terms do not affect the single-particle occupation, while $H_f$ conserves $n_\gamma=\langle\gamma^\dagger_\sigma\gamma_\sigma\rangle$ (SM, Sec.~\ref{sec:SM_rates}).
Consequently, the loss channel removes an occupied quasiparticle at rate $\Gamma_-n_\gamma$, whereas the gain channel fills an empty state at rate $\Gamma_+(1-n_\gamma)$, giving
\begin{equation}
\dot n_\gamma=\Gamma_+(1-n_\gamma)-\Gamma_-n_\gamma.
\label{eq:rate_equation_main}
\end{equation}
In the steady state, $\dot{n}_\gamma = 0$, and we obtain
\begin{equation}
n_\gamma^{\rm ss}=\frac{\Gamma_+}{\Gamma_++\Gamma_-},
\qquad
P_\gamma=1-2n_\gamma^{\rm ss}=\frac{\Gamma_- - \Gamma_+}{\Gamma_-+\Gamma_+}.
\label{eq:polarization_main}
\end{equation}
Both rates vanish as $1/\kappa$, reflecting the Zeno suppression of leakage, while their ratio remains finite and determines the steady occupation.
Thus $P_\gamma$ is a nonequilibrium occupation imbalance set by the reservoir rates rather than by a thermal filling factor, and inherits their strong dependence on $\phi$ and $\chi$.
The steady state is understood by first taking the long-time limit at large but finite $\kappa$ and only then sending $\kappa\to\infty$.

To evaluate the Josephson current in the dark sector, we employ 
$I_f=-2e\partial_\phi H_f$,
which, to leading order, gives
\begin{equation*}
I_{\rm ss}(\phi,\chi)=I_A(\phi,\chi)P_\gamma(\phi,\chi),
\end{equation*}
with
$I_A(\phi,\chi)=e{\Gamma^2\sin\eta\cos\eta}/{E_A}$,
recovering Eq.~\eqref{eq:intro_factorized}.
Here, the pair coherences generated by the interference terms are negligible to leading order in the secular regime $\Gamma_\pm\ll2E_A$ (SM, Sec.~\ref{sec:SM_rates}).
At fixed $\chi$, the two phases $\phi=-2\chi\pm\delta$ correspond to $\eta=\pm\delta/2$.
They have the same $E_A$ but opposite $I_A$, ensuring equal forward and backward critical-current magnitudes for the spectral channel.
The reservoir-selected occupation imbalance $P_\gamma$, however, need not be equal at these two phases.
Through $I_{\rm ss}=I_AP_\gamma$, this unequal weighting of opposite spectral currents can produce rectification (SM, Sec.~\ref{sec:SM_current}).
Within this approximation, rectification therefore arises entirely from the reservoir-selected occupation imbalance $P_\gamma$.
We quantify rectification by the signed diode efficiency
\begin{equation}
\eta_D(\chi)=\frac{I_c^+(\chi)-|I_c^-(\chi)|}
{I_c^+(\chi)+|I_c^-(\chi)|},
\label{eq:diode_efficiency_main}
\end{equation}
where $I_c^+$ and $|I_c^-|$ are the forward and backward critical-current magnitudes over one phase period.

\begin{figure}
\centering
\includegraphics[width=\columnwidth]{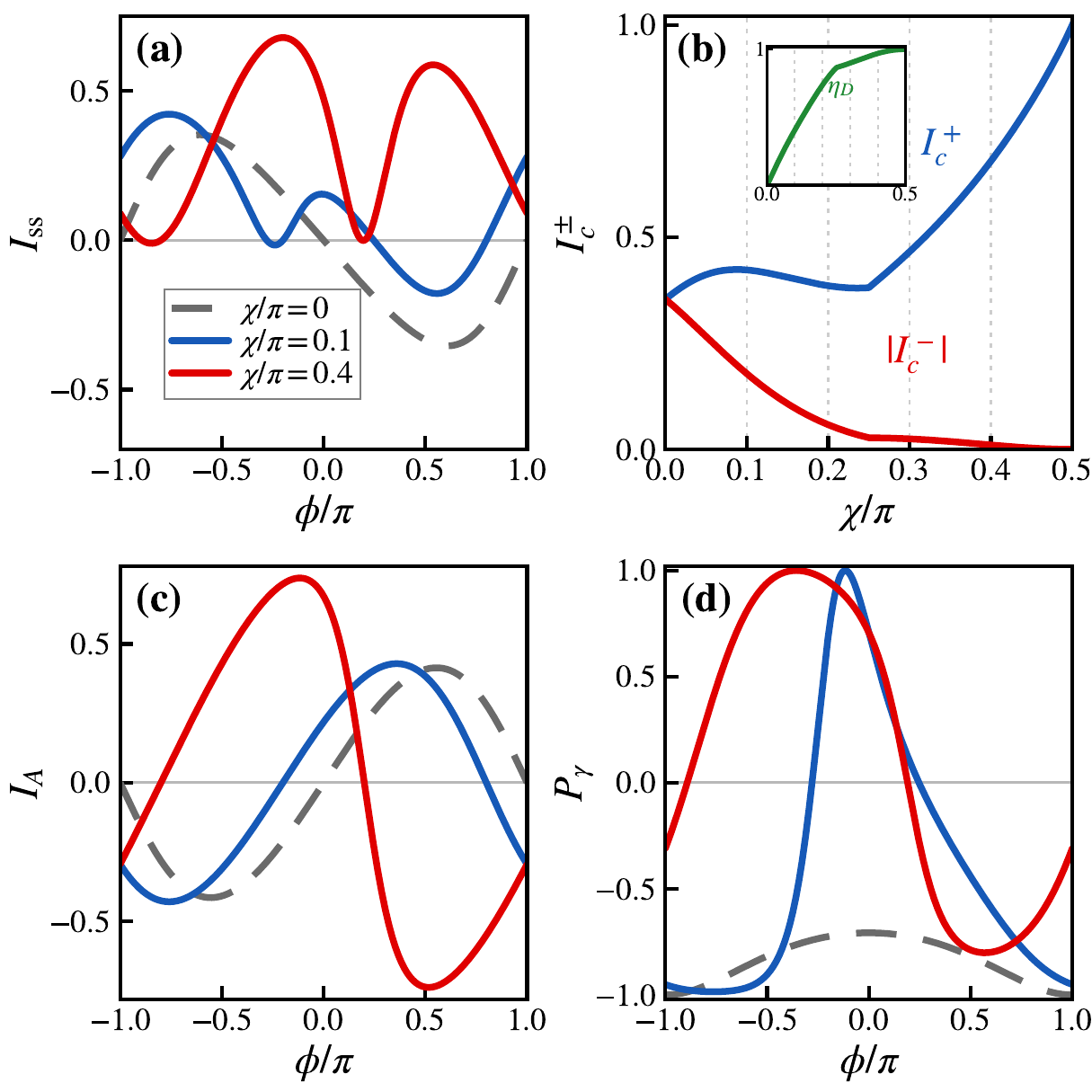}
\caption{
Occupation-reweighted Josephson diode.
(a) Steady-state current $I_{\rm ss}$ for $\chi/\pi=0$, $0.1$, and $0.4$.
(b) Forward and backward critical currents and the diode efficiency $\eta_D$ (inset) as functions of $\chi$.
The factorization in Eq.~\eqref{eq:intro_factorized} is resolved into
(c) the spectral Andreev current $I_A$ and
(d) the dark-channel occupation imbalance $P_\gamma$.
}
\label{fig:diode}
\end{figure}

\paragraph{Occupation-reweighted diode.---}
Figure~\ref{fig:diode} shows reservoir-induced Josephson rectification through occupation reweighting [Eq.~\eqref{eq:intro_factorized}].
We set $\Gamma=t=1$, which fixes the unit of energy.
At $\chi=0$, the CPR in panel (a) is reciprocal, whereas for the finite $\chi$ shown, the backward branch is strongly suppressed and the forward branch remains on the original current scale.
Panel (b) plots the corresponding critical currents and the diode efficiency, which approaches the ideal limit when $\chi$ increases.
Panels (c) and (d) reveal the origin of this asymmetry. 
The spectral current $I_A$ remains reciprocal for all $\chi$, whereas the occupation imbalance $P_\gamma$ varies strongly with phase and can become negative, reversing the sign of the steady-state current relative to $I_A$.
This occupation reweighting can produce both a nonzero phase-averaged current and a distortion of the CPR.

\begin{figure*}
\centering
\includegraphics[width=0.85\textwidth]{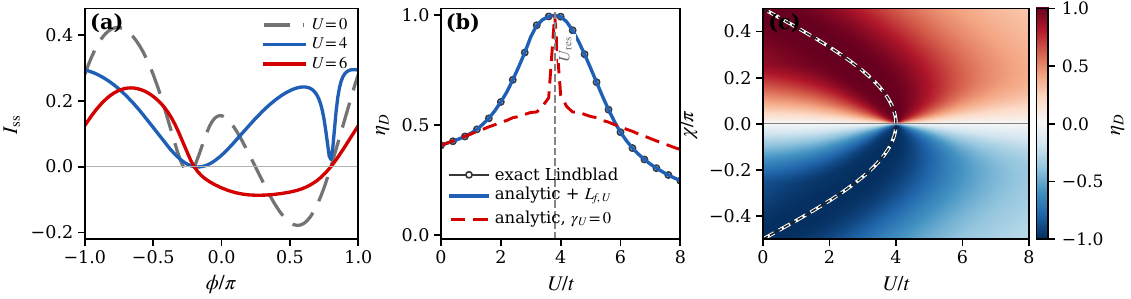}
\caption{
Interaction-assisted dark-pair resonance.
(a) Current-phase relations at
$U/t=0$, $4$, and $6$ for $\chi/\pi=0.1$ and $\kappa=300$.
(b) Diode efficiency $\eta_D$ versus $U/t$ at the same reservoir phase.
The full microscopic Lindblad result is compared with the interacting projected theory including the induced two-particle loss $L_{U,f}$ and with the corresponding theory without this channel, $\gamma_U=0$.
The vertical dashed line marks the dark-pair resonance
$U_{\rm res}=4t\cos\chi$.
(c) Signed diode efficiency $\eta_D$ of the interacting projected steady state in the $(U/t,\chi/\pi)$ plane for $\kappa=300$.
The dashed curve follows
$U_{\rm res}=4t\cos\chi$ [Eq.~\eqref{eq:interaction_resonance_main}].
Throughout, $t=\Gamma_L=\Gamma_R=1$.
}
\label{fig:interaction_resonance}
\end{figure*}

\paragraph{Interaction-assisted ideal diode.---}
Local Coulomb repulsion can enhance rectification by tuning a coherent dark-pair resonance and modifying the reservoir-selected occupations.
To see this, we add $H_U=U\sum_{j=L,R}\hat n_{j\uparrow}\hat n_{j\downarrow}$ to the original Hamiltonian $H$ and remain in the Zeno regime $\kappa\gg |t|,\Gamma,U$.
In the bright-dark basis, this local interaction can be decomposed as $H_U=H_{U,f}+H_{U,{\rm pair}}+H_{U,{\rm rem}}$, where $H_{U,f}=(U/2)\hat n_{f\uparrow}\hat n_{f\downarrow}$ acts within the dark sector and $H_{U,{\rm pair}}$ converts dark and bright pairs into one another.
The remainder acts only when bright states are already occupied and gives higher-order corrections.

The purely dark interaction, $H_{U,f}$, raises the energy of the doubly occupied dark state by $U/2$, changing its detuning from the empty state to $\delta_U=-2t\cos\chi+U/2$.
At $\delta_U=0$, these two states are resonantly mixed by the pairing term in $H_f$.
Coulomb repulsion thus tunes a coherent pair resonance at
\begin{equation}
U_{\rm res}(\chi) = 4t\cos\chi.
\label{eq:interaction_resonance_main}
\end{equation}
The steady current near this resonance also depends on the occupations selected by dissipation.
Here $H_{U,{\rm pair}}$ plays a distinct role: virtual conversion into a bright pair followed by its rapid decay removes a dark pair, producing the effective jump $L_{U,f}=\sqrt{\gamma_U}\,f_\downarrow f_\uparrow$ with rate $\gamma_U=U^2/(2\kappa)$.
Although this loss is suppressed as $\kappa$ increases, it competes with the one-body dissipative channels, whose rates also scale as $1/\kappa$.
It therefore remains essential for determining the steady-state occupations and current in the Zeno limit.
The full effective master equation, including the accompanying coherent correction to the dark pairing amplitude, is derived in the SM (Sec.~\ref{sec:interaction_benchmark}).

Figure~\ref{fig:interaction_resonance}(a) shows that tuning $U$ near the dark-pair resonance strongly suppresses the backward-current branch while retaining a finite forward supercurrent.
This suppression can be understood analytically at the dark-pair resonance.
Solving the interacting dark-sector master equation, including the induced pair loss, at $\delta_U=0$ yields the leading infinite-gap Zeno current (SM, Sec.~\ref{sec:interaction_benchmark}),
\begin{equation}
I_{\rm ss}^{\rm res}(\phi,\chi)
=
\frac{2et\Gamma^2\sin\chi\,\sin^2\eta}
{t^2+\Gamma^2\sin^2\eta}
+O(\kappa^{-1}).
\label{eq:resonant_current_main}
\end{equation}
For $\sin\chi\ne0$, the leading current has a fixed sign throughout the phase cycle, yielding ideal rectification at resonance.
Panel (b) benchmarks the effective theory by comparing $\eta_D$ extracted from the phase extrema of each steady-state CPR using Eq.~\eqref{eq:diode_efficiency_main}.
The black circles are obtained by numerically solving the full interacting two-dot Lindblad equation.
The blue solid curve uses the analytic steady-state current of the projected dark-sector theory, including $L_{U,f}$, whereas the red dashed curve sets $\gamma_U=0$ in the same theory (SM, Sec.~\ref{sec:interaction_benchmark}).
Thus, the coherent resonance sets the relevant interaction scale, while dissipative state selection determines the width and shape of the efficiency peak.
Panel (c) shows near-ideal rectification along $U_{\rm res}(\chi)$.
The resonance position is unchanged under $\chi\to-\chi$, while the diode polarity reverses: interactions tune the pair resonance, and the reservoir phase selects the current direction.

\begin{figure}
\centering
\includegraphics[width=1.0\linewidth]{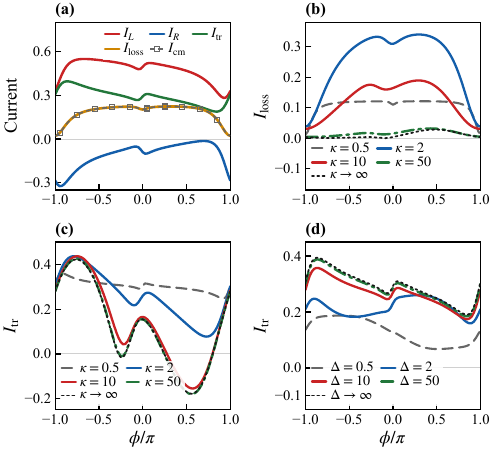}
\caption{
Currents beyond the infinite-gap Zeno limits at $\chi/\pi=0.1$.
(a) Various currents versus $\phi$ at $\kappa=\Delta=1$. $I_{\rm cm}=I_{\rm loss}$ verifies steady-state continuity.
Panels (b) and (c) show the current-phase relations of $I_{\rm loss}$ and $I_{\rm tr}$, respectively, at $\Delta=1$ for different values of $\kappa$.
Strong dissipation depletes the bright sector, reducing leakage while the dark channel retains a finite supercurrent.
(d) $I_{\rm tr}(\phi)$ for various $\Delta$ at $\kappa=1$.
}
\label{fig:robustness}
\end{figure}

\paragraph{Beyond the infinite-gap Zeno limits.---}
The results above rely on two controlled limits:
the Zeno limit, in which the bright sector is eliminated, and the infinite-gap limit, in which the superconducting leads induce instantaneous pairing.
We now relax both approximations using a Keldysh-Lindblad Green's function formulation~\cite{Sieberer2016Keldysh,Stefanucci2024KadanoffBaym} (SM, Sec.~\ref{sec:SM_keldysh}).

At finite dissipation, the bright sector remains dynamically active and the two lead currents are no longer constrained to be equal and opposite.
We take both $I_L$ and $I_R$ to be positive for charge flow from the corresponding lead into the junction.
Steady-state charge conservation then gives
$I_L+I_R=I_{\rm loss,tot}$,
where $I_{\rm loss,tot}$ is the total reservoir outflow.
It is therefore useful to separate the lead currents into transport and common-mode components,
\begin{equation}
I_{\rm tr}=(I_L-I_R)/2,\qquad
I_{\rm cm}=(I_L+I_R)/2.
\label{eq:transport_common_mode}
\end{equation}
Equivalently, $I_L=I_{\rm cm}+I_{\rm tr}$ and $I_R=I_{\rm cm}-I_{\rm tr}$: the transport component contributes with opposite signs at the two terminals, while the common-mode component contributes equally to both.
The factors $1/2$ normalize the decomposition so that a lossless current $I_L=-I_R=J$ gives $I_{\rm tr}=J$.
Steady-state continuity gives $I_{\rm cm}=I_{\rm loss,tot}/2 := I_{\rm loss}$.

Figure~\ref{fig:robustness}~(a) verifies $I_{\rm cm}=I_{\rm loss}$ at $\kappa=\Delta=1$.
In particular, $I_L$ and $I_R$ are visibly not mirror images, while $I_{\rm cm}$ coincides with the spin-averaged reservoir outflow $I_{\rm loss}$, shown by open squares.
Because either lead current mixes these two components, we characterize the diode response by the transport current $I_{\rm tr}$ rather than by $I_L$ or $I_R$ separately.
Panels (b) and (c) show $I_{\rm loss}$ and $I_{\rm tr}$ for various dissipation strengths $\kappa$.
The leakage first grows with $\kappa$ and then decreases at strong dissipation.
This turnover reflects the dependence of the actual loss on both the decay rate and the bright-sector occupation: $I_{\rm loss}\propto\kappa n_c$, where $n_c=\sum_\sigma\langle c_\sigma^\dagger c_\sigma\rangle$.
Weak loss increases the escape rate without strongly suppressing $n_c$.
In the infinite-gap Zeno theory, however, a coherent coupling of strength $g$ into the bright sector produces an amplitude of order $g/\kappa$ and hence an occupation of order $(g/\kappa)^2$.
The resulting leakage scales as $\kappa(g/\kappa)^2=g^2/\kappa$: stronger damping reduces the actual loss by suppressing population of the lossy sector.
This continuous quantum Zeno suppression has also been reported in ultracold molecular gases~\cite{Syassen2008Science}.
The transport current in panel~(c) can remain finite and exhibit diode behavior because the dark channel retains coherent pairing.
Finally, Fig.~\ref{fig:robustness}(d) examines the effect of a finite superconducting gap at $\kappa=1$.
Retardation and coupling to the quasiparticle continuum quantitatively reshape the CPR, but the branch asymmetry remains visible down to $\Delta\sim t$ and evolves smoothly toward the instantaneous-pairing result as $\Delta$ increases.
The diode response is therefore not an artifact of the infinite-gap Zeno limit.

\paragraph{Conclusions.---}
We have demonstrated a reservoir-engineered Josephson diode in a double-dot junction with symmetric coherent couplings and polarity controlled by the reservoir phase.
In the infinite-gap Zeno regime, the leading secular current factorizes as $I_{\rm ss}=I_A P_\gamma$, identifying reservoir-selected occupation reweighting of a reciprocal spectral current as the source of rectification.
Furthermore, Coulomb repulsion tunes a dark-pair resonance that enables near-ideal rectification, with induced two-particle loss essential to the quantitative description of the steady current.
Beyond the infinite-gap Zeno regime, Keldysh-Lindblad calculations show that nonreciprocal transport persists at finite loss strength and superconducting gap.
In this case, the loss current depends on both the decay rate and the bright-sector occupation.
At strong dissipation, suppression of this occupation can outweigh the increase in the decay rate, reducing leakage, while the dark channel continues to carry a finite Josephson current.

\begin{acknowledgments}
This work is supported by the National Natural Science Foundation of China (Grant No. 12504288) and the Science and Technology Commission of Shanghai Municipality (Grant No. 25ZR1402550).
\end{acknowledgments}

\clearpage
\onecolumngrid
\setcounter{secnumdepth}{2}
\renewcommand{\thesection}{S\arabic{section}}
\renewcommand{\thesubsection}{\Alph{subsection}}
\renewcommand{\theequation}{S\arabic{equation}}
\renewcommand{\thefigure}{S\arabic{figure}}
\renewcommand{\thetable}{S\arabic{table}}
\setcounter{section}{0}
\setcounter{equation}{0}
\setcounter{figure}{0}
\setcounter{table}{0}

\section*{Supplemental Material: Reservoir-Engineered Josephson Diode in a Dark Andreev Channel}

\section{Overview}
\label{sec:SM_overview}

This Supplemental Material provides the derivations and numerical benchmarks supporting the main text.
Section~\ref{sec:SM_projected_theory} works in two controlled limits, an infinite superconducting gap and a loss rate $\kappa$ larger than any other dynamical scale (quantum Zeno limit), so that the problem remains analytically tractable. 
Starting from a lossy auxiliary mode, we derive the phase-textured nonlocal Lindblad operator, transform to the bright-dark basis it selects, and eliminate the bright mode. A secular reduction then yields the factorized steady-state current for the noninteracting junction.
We then examine the robustness against pairing asymmetry and hopping variations, and derive an effective model in the presence of local Coulomb interaction that incorporates the dark-pair resonance and induced two-particle loss.
Section~\ref{sec:SM_keldysh} goes beyond the infinite-gap Zeno limit and treats the junction at finite loss strength and superconducting gap using the Keldysh-Lindblad formalism.
The linear Lindblad operators are recast as Keldysh self-energies, allowing the lead currents, reservoir outflow, and transport response to be evaluated within nonequilibrium Green's function theory.
Unless stated otherwise, symbols have the same meaning as in the main text, and we set $\hbar =1$.

\section{Zeno effective theory in the infinite-gap limit}
\label{sec:SM_projected_theory}

The analytic theory in this section uses two controlled limits: (i) the infinite-gap limit, where the superconducting leads reduce to static pairing amplitudes on the dots, and (ii) the Zeno limit, where the loss rate $\kappa$ is the largest dynamical scale, allowing the bright mode to be eliminated. 
Section~\ref{sec:SM_keldysh} relaxes both assumptions.

\subsection{Auxiliary-mode construction of the nonlocal loss}
\label{sec:SM_auxiliary}

\begin{figure}
\centering
\resizebox{0.40\linewidth}{!}{%
\begin{tikzpicture}[
  >={Stealth[length=2.4mm,width=1.9mm]},
  dot/.style   ={circle,draw=blue!55!black,thick,fill=blue!8,minimum size=9.5mm,inner sep=0pt},
  aux/.style   ={circle,draw=red!65!black,thick,fill=red!8,minimum size=9.5mm,inner sep=0pt},
  drain/.style ={rectangle,draw=black!70,thick,fill=black!5,rounded corners=1pt,
                 minimum width=10mm,minimum height=6.5mm,inner sep=1pt},
  cpl/.style   ={->,thick,blue!55!black},
  loss/.style  ={->,very thick,red!70!black},
  slbl/.style  ={font=\scriptsize}
]

\node[slbl,black!70] at (1.5,3.02) {lossy auxiliary mode};

\node[dot] (dL) at (0,0)      {$d_{L\sigma}$};
\node[dot] (dR) at (3.0,0)    {$d_{R\sigma}$};
\node[aux] (a)  at (1.5,2.15) {$a_{\sigma}$};
\node[drain] (drn) at (4.60,2.15) {\footnotesize drain};

\draw[<->,semithick,black!45] (dL) -- node[below,slbl,black!60,yshift=-1pt]{$H$} (dR);
\draw[cpl] (dL) -- node[slbl,pos=0.55,left,xshift=1pt]{$\dfrac{ge^{i\chi}}{\sqrt{2}}$} (a);
\draw[cpl] (dR) -- node[slbl,pos=0.55,right,xshift=-1pt]{$\dfrac{g}{\sqrt{2}}$} (a);
\draw[->,dashed,semithick,black!55] (2.02,0.85) arc (0:290:0.52);
\node[slbl,black!65] at (1.5,0.85) {$\chi$};
\draw[loss] (a) -- node[above,slbl,red!70!black]{$\kappa_a$} (drn);

\end{tikzpicture}}
\caption{
Auxiliary-mode construction of the nonlocal loss. 
For each spin, the two dot modes couple to the auxiliary mode with equal magnitude $|g|/\sqrt{2}$ and relative phase $\chi$. 
The auxiliary mode is emptied into a Markovian drain at the population-decay rate $\kappa_a$, and $H$ denotes the coherent double-dot Hamiltonian.
}
\label{fig:SM_aux_schematic}
\end{figure}
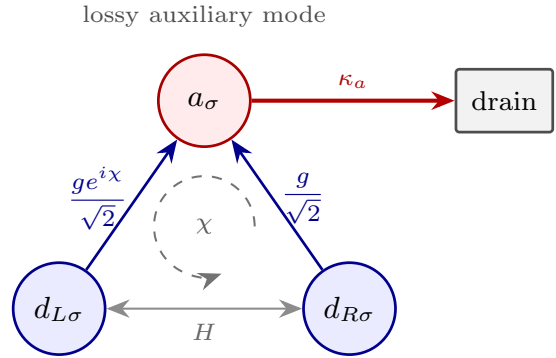

The nonlocal loss channel in Eq.~\eqref{eq:nonlocal_jump} of the main text can be realized by coupling a common reservoir to a coherent superposition of local modes~\cite{Dicke1954,Lidar1998DFS,Poyatos1996PRL}.
Here we derive it from the lossy auxiliary-mode setup of Fig.~\ref{fig:SM_aux_schematic}: for each spin $\sigma$, an auxiliary fermionic mode $a_\sigma$ tunnels to both dot modes with equal magnitude and a relative phase $\chi$ set by the magnetic field,
\begin{equation}
H_{\rm aux}
=
\sum_\sigma
\left[
\delta_a a^\dagger_\sigma a_\sigma
+
\frac{g}{\sqrt{2}}a^\dagger_\sigma
\left(
d_{R\sigma}+e^{i\chi}d_{L\sigma}
\right)
+
\frac{g^*}{\sqrt{2}}
\left(
d^\dagger_{R\sigma}+e^{-i\chi}d^\dagger_{L\sigma}
\right)
a_\sigma
\right].
\label{eq:SM_Haux}
\end{equation}
Here $\delta_a$ is the auxiliary-mode on-site energy, $g$ the tunneling amplitude, and $H$ the coherent double-dot Hamiltonian. 
With the normalized dot orbital
\begin{equation}
c_\sigma
=
\frac{d_{R\sigma}+e^{i\chi}d_{L\sigma}}{\sqrt{2}},
\label{eq:SM_bright_def}
\end{equation}
Eq.~\eqref{eq:SM_Haux} reduces to
\begin{equation}
H_{\rm aux}
=
\sum_\sigma
\left[
\delta_a a^\dagger_\sigma a_\sigma
+\left(g a^\dagger_\sigma c_\sigma+{\rm H.c.}\right)
\right].
\label{eq:SM_Haux_B}
\end{equation}
Thus $a_\sigma$ couples only to $c_\sigma$ and the orthogonal mode $f_\sigma=(-e^{-i\chi}d_{R\sigma}+d_{L\sigma})/\sqrt{2}$ is left untouched. 
The auxiliary mode is emptied into a Markovian drain through the jump operator
\begin{equation}
L_{a\sigma}=\sqrt{\kappa_a}a_\sigma,
\label{eq:SM_La}
\end{equation}
with $\kappa_a$ its population-decay rate. 
The density matrix $\rho$ of the dots and auxiliary mode obeys
\begin{equation}
\dot\rho
=
\underbrace{
-i[H+H_{\rm aux},\rho]
+
\sum_\sigma \mathcal{D}[L_{a\sigma}]\rho
}_{:=\mathcal{L} \rho},
\quad
\text{with}
\quad
\mathcal{D}[L]\rho
=
L\rho L^\dagger-
\frac{1}{2}\{L^\dagger L,\rho\}.
\label{eq:SM_master_aux}
\end{equation}
Here, $[A,B]=AB-BA$, $\{A,B\}=AB+BA$.

To eliminate $a_\sigma$ and obtain a reduced equation for the dots~\cite{Reiter2012PRA}, we first evaluate the bare retarded Green's function of the auxiliary mode,
\begin{equation}
\mathsf{g}_a^R(t)
=
-i\theta(t)
\left\langle
\left\{a_\sigma(t),a_\sigma^\dagger(0)\right\}
\right\rangle,
\label{eq:SM_aux_gR_time}
\end{equation}
where $\theta(t)$ is the Heaviside step function and ``bare'' means $g=0$ ($\delta_a$ and $\kappa_a$ retained). 
The expectation value follows from the quantum regression theorem~\cite{Lax1963,GardinerZoller2004},
\begin{equation}
X(t)
\equiv
\left\langle
\left\{a_\sigma(t),a_\sigma^\dagger(0)\right\}
\right\rangle
=
{\rm Tr}
\left[
a_\sigma
e^{\mathcal{L}_a t}
\left(
a_\sigma^\dagger\rho_{\rm ss}
+
\rho_{\rm ss}a_\sigma^\dagger
\right)
\right],
\end{equation}
where $\mathcal{L}_a$ is the generator of Eq.~\eqref{eq:SM_master_aux} at $g=0$, built from $H_a=\delta_a\sum_\sigma a_\sigma^\dagger a_\sigma$ and $L_{a\sigma}$, and $\rho_{\rm ss}$ is its stationary state. 
Transferring the generator onto the left argument gives $\dot X={\rm Tr}[\mathcal{L}_a^\dagger(a_\sigma)\cdots]$. The opposite-spin dissipator drops out inside this vacuum trace, leaving
\begin{equation}
i\left[H_a,a_\sigma\right]
+
\kappa_a
\left(
a_\sigma^\dagger a_\sigma a_\sigma
-
\frac{1}{2}
\left\{
a_\sigma^\dagger a_\sigma,a_\sigma
\right\}
\right)
=
-\left(
i\delta_a+\frac{\kappa_a}{2}
\right)a_\sigma.
\end{equation}
Hence $\dot X=-(i\delta_a+\kappa_a/2)X$ with $X(0)=1$ from the canonical anticommutation relation, so that $\mathsf{g}_a^R(t)=-i\theta(t)e^{-(i\delta_a+\kappa_a/2)t}$, and Fourier transformation yields
\begin{equation}
\mathsf{g}_a^R(\omega)
=
\frac{1}{\omega-\delta_a+i\kappa_a/2}.
\label{eq:SM_aux_gR}
\end{equation}
The retarded self-energy induced on $c_\sigma$ is
\begin{equation}
\Sigma_c^R(\omega)
=
g^* \mathsf{g}_a^R(\omega)g
=
\frac{|g|^2}{\omega-\delta_a+i\kappa_a/2}.
\label{eq:SM_aux_selfenergy_exact}
\end{equation}
Its real part shifts the level and its imaginary part broadens it.

The local Markov limit requires $\mathsf{g}_a^R(\omega)$ to be flat over the frequency window explored by the dots,
\begin{equation}
\max\left\{|\omega|,\Omega_{\rm dot},|g|\right\}
\ll
\Omega_a
\equiv
\sqrt{(\kappa_a/2)^2+\delta_a^2}.
\label{eq:SM_aux_regime}
\end{equation}
Here $\omega$ is the probe frequency, $\Omega_{\rm dot}$ the largest dynamical scale of the dots, and the third entry makes the replacement self-consistent: elimination dresses $c_\sigma$ with the shift and width $|\Sigma_c^R(0)|=|g|^2/\Omega_a$. 
The auxiliary response then varies little over the relevant dot-frequency window, so Eq.~\eqref{eq:SM_aux_selfenergy_exact} may be replaced by its zero-frequency value, $\Sigma_c^R(\omega)=\Sigma_c^R(0)[1+O(\omega/\Omega_a)]$:
\begin{equation}
\Sigma_c^R(0)
=
\varepsilon_{\rm LS}
-
\frac{i\kappa_{\rm eff}}{2},
\quad
\text{with}
\quad
\kappa_{\rm eff}
=
\frac{|g|^2\kappa_a}{(\kappa_a/2)^2+\delta_a^2}
\quad
\text{and}
\quad
\varepsilon_{\rm LS}
=
-\frac{|g|^2\delta_a}{(\kappa_a/2)^2+\delta_a^2}.
\label{eq:SM_Leff_B}
\end{equation}
Here, the real part gives the coherent Lamb shift (LS) of $c_\sigma$, while the imaginary part gives its amplitude-decay rate $\kappa_{\rm eff}/2$.

The retarded self-energy does not determine whether the reservoir injects particles. 
That information is carried by the lesser component. 
Applying the adjoint generator to the occupation $\hat n_{a\sigma}=a_\sigma^\dagger a_\sigma$ gives
\begin{equation}
\frac{d}{dt}\langle\hat n_{a\sigma}\rangle
=
\left\langle
\mathcal{L}_a^\dagger(\hat n_{a\sigma})
\right\rangle
=
-\kappa_a\langle\hat n_{a\sigma}\rangle.
\end{equation}
With no gain channel, the auxiliary mode relaxes uniquely to the vacuum, $\langle\hat n_{a\sigma}\rangle_{\rm ss}=0$ ($L_{a\sigma}|0_a\rangle=0$). 
The regression theorem then gives $\langle a_\sigma^\dagger(t')a_\sigma(t)\rangle_{\rm ss} =e^{-(i\delta_a+\kappa_a/2)\tau}\langle\hat n_{a\sigma}\rangle_{\rm ss}=0$ for $\tau=t-t'\geq0$, and likewise for $t<t'$ by complex conjugation.  
Consequently,
\begin{equation}
\mathsf{g}_a^<(t,t')
=
i\left\langle a_\sigma^\dagger(t')a_\sigma(t)\right\rangle_{\rm ss}
=0,
\qquad
\Sigma_c^<(\omega)
=
g^* \mathsf{g}_a^<(\omega)g
=0.
\label{eq:SM_aux_lesser}
\end{equation}
The eliminated auxiliary mode thus supplies no incoming-particle source in the $c_\sigma$ channel.

The self-energies fix the level shift and linewidth but not the jump term required for a trace-preserving master equation. 
To derive it, we treat the double dot as the slow system and the rapidly relaxing auxiliary mode as its environment, with coupling from Eq.~\eqref{eq:SM_Haux_B},
\begin{equation}
V
=
\sum_\sigma
\left(
g^*c_\sigma^\dagger a_\sigma
+
g a_\sigma^\dagger c_\sigma
\right).
\label{eq:SM_aux_bath_force}
\end{equation}
Since $\langle a_\sigma\rangle_{\rm ss}=0$, the leading correction to the reduced dot state $\rho_{\rm dot}=\operatorname{Tr}_a\rho$ is second order in $V$: two coupling vertices separated by $\tau$, during which the auxiliary mode evolves with $\mathcal{L}_a$ while the slow dot state is held fixed,
\begin{equation}
\left.\dot\rho_{\rm dot}\right|_a
=
\int_0^\infty d\tau\,
\operatorname{Tr}_a
\left\{
\mathcal{L}_V
e^{\mathcal{L}_a\tau}
\mathcal{L}_V
\left[
\rho_{\rm dot}(t)\otimes\rho_a^{\rm ss}
\right]
\right\},
\label{eq:SM_aux_TCL_generator}
\end{equation}
where $\mathcal{L}_V X=-i[V,X]$ and $\rho_a^{\rm ss}=|0_a\rangle\langle0_a|$. 
The same vacuum regression argument as in Eqs.~\eqref{eq:SM_aux_gR_time}-\eqref{eq:SM_aux_gR} gives exponential decay at rate $\kappa_a/2$ and oscillation frequency $\delta_a$. The equal-time values follow from the canonical anticommutation relation and $\langle\hat n_{a\sigma}\rangle_{\rm ss}=0$. Thus, for $\tau\geq0$,
\begin{equation}
\left\langle
a_\sigma(\tau)a_{\sigma'}^\dagger(0)
\right\rangle_{\rm ss}
=
\delta_{\sigma\sigma'}
e^{-(i\delta_a+\kappa_a/2)\tau},
\qquad
\left\langle
a_{\sigma'}^\dagger(0)a_\sigma(\tau)
\right\rangle_{\rm ss}
=0.
\end{equation}
Expanding the double commutator in Eq.~\eqref{eq:SM_aux_TCL_generator} generates four orderings, distinguished by whether each vertex acts from the left or right of $\rho_{\rm dot}\otimes\rho_a^{\rm ss}$. 
Each carries one $a_\sigma$ and one $a_\sigma^\dagger$, so the auxiliary trace factorizes onto the correlators above. 
Only the greater correlator survives, every ordering carries the weight $|g|^2$, and vertices acting from the right carry the conjugate kernel $e^{(i\delta_a-\kappa_a/2)\tau}$.

With $\rho_{\rm dot}$ fixed over the correlation time, the dot operators factor out of the $\tau$ integral in Eq.~\eqref{eq:SM_aux_TCL_generator} and only the one-sided kernel enters:
\begin{equation}
K
\equiv
|g|^2\int_0^\infty d\tau\,
e^{-(i\delta_a+\kappa_a/2)\tau}
=
\frac{|g|^2}{\kappa_a/2+i\delta_a}
=
i\Sigma_c^R(0)
=
\frac{\kappa_{\rm eff}}{2}
+i\varepsilon_{\rm LS}.
\label{eq:SM_aux_K_kernel}
\end{equation}
The conjugate kernel gives $K^*$, so the four orderings collapse onto the pair $K$ and $K^*$. 
The one-sided restriction keeps $K$ complex, determining the population-decay rate $\kappa_{\rm eff}=2{\rm Re}\,K$ and Lamb shift $\varepsilon_{\rm LS}={\rm Im}\,K$ together, whereas the full-line integral $K+K^*=\kappa_{\rm eff}$ would discard the shift.
Substituting into Eq.~\eqref{eq:SM_aux_TCL_generator} gives
\begin{equation}
\left.\dot\rho_{\rm dot}\right|_a
=
\sum_\sigma
\left[
K
\left(
c_\sigma\rho_{\rm dot}c_\sigma^\dagger
-c_\sigma^\dagger c_\sigma\rho_{\rm dot}
\right)
+
K^*
\left(
c_\sigma\rho_{\rm dot}c_\sigma^\dagger
-\rho_{\rm dot}c_\sigma^\dagger c_\sigma
\right)
\right].
\label{eq:SM_aux_master_from_kernel}
\end{equation}
Using $K=\kappa_{\rm eff}/2+i\varepsilon_{\rm LS}$ and grouping the terms into a commutator and a dissipator yields
\begin{equation}
\left.\dot\rho_{\rm dot}\right|_a
=
-i[H_{\rm LS},\rho_{\rm dot}]
+
\kappa_{\rm eff}
\sum_\sigma
\left(
c_\sigma\rho_{\rm dot}c_\sigma^\dagger
-\frac{1}{2}c_\sigma^\dagger c_\sigma\rho_{\rm dot}
-\frac{1}{2}\rho_{\rm dot}c_\sigma^\dagger c_\sigma
\right),
\qquad
H_{\rm LS}
=
\varepsilon_{\rm LS}
\sum_\sigma c_\sigma^\dagger c_\sigma.
\label{eq:SM_aux_master_explicit}
\end{equation}
Equation~\eqref{eq:SM_aux_master_explicit} displays the recycling term $c_\sigma\rho_{\rm dot}c_\sigma^\dagger$ that the retarded self-energy alone does not fix, and identifies the effective jump operator
\begin{equation}
L_\sigma^{\rm eff}
=
\sqrt{\kappa_{\rm eff}}c_\sigma.
\label{eq:SM_effective_aux_jump}
\end{equation}
The leading local master equation for the double dot is therefore
\begin{equation}
\dot\rho_{\rm dot}
=
-i[H+H_{\rm LS},\rho_{\rm dot}]
+
\sum_\sigma \mathcal{D}[L_\sigma^{\rm eff}]\rho_{\rm dot}.
\label{eq:SM_eff_master_dot_eq}
\end{equation}
Corrections arise from the frequency dependence of Eq.~\eqref{eq:SM_aux_selfenergy_exact} and are smaller by the ratios $|\omega|/\Omega_a$ and $\Omega_{\rm dot}/\Omega_a$.

For finite $\delta_a$ the elimination produces both the coherent shift and the loss. 
In the resonant case used in the main text, $\delta_a=0$, the shift vanishes and
\begin{equation*}
\kappa
\equiv
\kappa_{\rm eff}
=
\frac{4|g|^2}{\kappa_a}.
\end{equation*}
Substituting Eq.~\eqref{eq:SM_bright_def} into Eq.~\eqref{eq:SM_effective_aux_jump} gives the nonlocal Lindblad operator used throughout the main text,
\begin{equation}
L_\sigma
=
\sqrt{\frac{\kappa}{2}}
\left(
d_{R\sigma}+e^{i\chi}d_{L\sigma}
\right).
\label{eq:SM_nonlocal_jump}
\end{equation}

\subsection{Gauge convention and bright-dark basis}
\label{sec:SM_bright_dark}

We use the infinite-gap double-dot Hamiltonian of the main text, in which the leads have been replaced by static pairing amplitudes on the dots.  
Section~\ref{sec:SM_KL_nambu} obtains this replacement from the lead self-energy and shows where it fails at finite gap.  
The Hamiltonian is
\begin{equation}
H
=
H_t+H_\Delta,
\label{eq:SM_Hsys}
\end{equation}
with
\begin{subequations}
\begin{align}
H_t
&=
t\sum_\sigma
\left(
d^\dagger_{L\sigma}d_{R\sigma}
+
d^\dagger_{R\sigma}d_{L\sigma}
\right),
\label{eq:SM_Ht}
\\
H_\Delta
&=
\Gamma_L e^{i\phi/2}
d^\dagger_{L\uparrow}d^\dagger_{L\downarrow}
+
\Gamma_R e^{-i\phi/2}
d^\dagger_{R\uparrow}d^\dagger_{R\downarrow}
+
{\rm H.c.}
\label{eq:SM_HDelta}
\end{align}
\end{subequations}
Here $t$ is the interdot hopping (unrelated to the time argument $t$ of the Green's functions in Sec.~\ref{sec:SM_auxiliary}) and $\Gamma_{L,R}$ are the induced pairing amplitudes, equal to the normal-state hybridization widths of Sec.~\ref{sec:SM_KL_nambu} in the infinite-gap limit.  
The dot on-site energies are set to zero throughout this section, and are kept general in Sec.~\ref{sec:SM_KL_nambu}.  
We work in the symmetric gauge
\begin{equation}
\phi_L=\frac{\phi}{2},
\qquad
\phi_R=-\frac{\phi}{2},
\label{eq:SM_symmetric_gauge}
\end{equation}
which treats the two superconducting leads on equal footing.

The phases assigned separately to the Hamiltonian and the Lindblad operator depend on the local phase convention of the dot fields.  
Only gauge-invariant combinations can enter observables.  
To identify them, temporarily write the hopping amplitude as $t e^{i\theta_t}$, with Eq.~\eqref{eq:SM_Ht} recovered at $\theta_t=0$.  
Under the local rephasing $d_{j\sigma}\to e^{i\theta_j}d_{j\sigma}$, and defining $\delta=\theta_L-\theta_R$, the phases transform as
\begin{equation*}
\begin{aligned}
\theta_t&\to\theta_t-\delta,
&
\phi_L&\to\phi_L-2\theta_L,
\\
\phi_R&\to\phi_R-2\theta_R,
&
\chi&\to\chi+\delta.
\end{aligned}
\end{equation*}
The pair phases acquire twice the local fermion phase, while the overall phase of $L_\sigma$ is immaterial because $\mathcal{D}[e^{i\alpha}L_\sigma]=\mathcal{D}[L_\sigma]$.

Two independent gauge-invariant combinations may therefore be chosen as
\begin{equation*}
\theta_t+\chi,
\qquad
\frac{\phi_L-\phi_R}{2}+\chi.
\end{equation*}
In the convention of Eqs.~\eqref{eq:SM_Ht}, \eqref{eq:SM_symmetric_gauge}, and \eqref{eq:SM_nonlocal_jump}, these reduce to $\chi$ and
\begin{equation}
\eta
\equiv
\frac{\phi_L-\phi_R}{2}+\chi
=
\frac{\phi}{2}+\chi.
\label{eq:SM_eta}
\end{equation}
The factor $1/2$ reflects that a local pair operator carries twice the phase of a single fermion.  
In this gauge, all coefficients obtained below can be expressed in terms of $\chi$ and $\eta$.

We now pass to the single-particle basis selected by the reservoir.  
The nonlocal Lindblad operator singles out the combination $c_\sigma$ of Eq.~\eqref{eq:SM_bright_def}.  
Completing it with its orthogonal partner gives
\begin{equation}
c_\sigma
=
\frac{
d_{R\sigma}
+
e^{i\chi}d_{L\sigma}
}{\sqrt{2}},
\qquad
f_\sigma
=
\frac{
-e^{-i\chi}d_{R\sigma}
+
d_{L\sigma}
}{\sqrt{2}}.
\end{equation}
The transformation is unitary and therefore preserves the fermionic anticommutation relations.  
Its inverse is
\begin{equation}
d_{R\sigma}
=
\frac{
c_\sigma
-
e^{i\chi}f_\sigma
}{\sqrt{2}},
\qquad
d_{L\sigma}
=
\frac{
e^{-i\chi}c_\sigma
+
f_\sigma
}{\sqrt{2}}.
\label{eq:SM_inverse_bright_dark}
\end{equation}
In this basis, Eq.~\eqref{eq:SM_nonlocal_jump} becomes $L_\sigma=\sqrt{\kappa}c_\sigma$. 
The dissipator is therefore diagonal in this basis: $c_\sigma$ is bright, whereas $f_\sigma$ is dark under the bare loss.  
The coherent Hamiltonian is not generally diagonal here, and its bright-dark couplings are derived next.

\subsection{Hamiltonian in the bright-dark basis}
\label{sec:SM_H_decomposition}

To express the coherent Hamiltonian in the bright-dark basis, we insert the inverse transformation of Eq.~\eqref{eq:SM_inverse_bright_dark} into Eq.~\eqref{eq:SM_Hsys}. 
We use $\Gamma_L=\Gamma_R\equiv\Gamma$ unless stated otherwise.
Pairing asymmetry, $\Gamma_L\neq\Gamma_R$, is examined separately in Sec.~\ref{sec:SM_asymmetry_robustness}.
We keep $t$ and $\Gamma$ independent and denote their ratio by $r\equiv t/\Gamma$. 
The hopping and pairing terms become
\begin{subequations}\label{eq:SM_H_bd}
\begin{equation}
H_t
=
t\cos\chi
\sum_\sigma
\left(
c^\dagger_\sigma c_\sigma
-
f^\dagger_\sigma f_\sigma
\right)
+
t\sum_\sigma
\left(
-i e^{i\chi}\sin\chi c^\dagger_\sigma f_\sigma
+
i e^{-i\chi}\sin\chi f^\dagger_\sigma c_\sigma
\right),
\label{eq:SM_Ht_bd}
\end{equation}
\begin{equation}
H_\Delta
=
\Gamma e^{i\chi}\cos\eta c^\dagger_\uparrow c^\dagger_\downarrow
+
\Gamma e^{-i\chi}\cos\eta f^\dagger_\uparrow f^\dagger_\downarrow
+
i\Gamma\sin\eta
\left(
c^\dagger_\uparrow f^\dagger_\downarrow
-
c^\dagger_\downarrow f^\dagger_\uparrow
\right)
+
{\rm H.c.}
\label{eq:SM_HDelta_bd}
\end{equation}
\end{subequations}
After rearranging the terms in Eq.~\eqref{eq:SM_H_bd}, the system Hamiltonian becomes
\begin{equation}
H
=
H_c+H_f+H_{cf},
\label{eq:SM_H_decomp}
\end{equation}
where the bright and dark sector Hamiltonians and their coupling read
\begin{subequations}
\begin{align}
H_c
&=
t\cos\chi
\sum_\sigma
c^\dagger_\sigma c_\sigma
+
\Gamma e^{i\chi}\cos\eta c^\dagger_\uparrow c^\dagger_\downarrow
+
\Gamma e^{-i\chi}\cos\eta c_\downarrow c_\uparrow,
\label{eq:SM_Hc}
\\
H_f
&=
-t\cos\chi
\sum_\sigma
f^\dagger_\sigma f_\sigma
+
\Gamma e^{-i\chi}\cos\eta f^\dagger_\uparrow f^\dagger_\downarrow
+
\Gamma e^{i\chi}\cos\eta f_\downarrow f_\uparrow,
\label{eq:SM_Hf}
\\
H_{cf} &=
-it e^{i\chi}\sin\chi
\left(
c^\dagger_\uparrow f_\uparrow
+
c^\dagger_\downarrow f_\downarrow
\right)
+
i\Gamma\sin\eta
\left(
c^\dagger_\uparrow f^\dagger_\downarrow
-
c^\dagger_\downarrow f^\dagger_\uparrow
\right)
+ {\rm H.c.}
\label{eq:SM_Hcf}
\end{align}
\end{subequations}

In Eq.~\eqref{eq:SM_H_decomp}, $H_f$ contains the dark Andreev channel retained in the Zeno limit and $H_c$ the short-lived bright sector. 
Their coupling $H_{cf}$ contains a normal conversion vertex proportional to $t\sin\chi$ and an anomalous bright-dark pair-creation vertex proportional to $\Gamma\sin\eta$. 
The next subsection eliminates the bright mode and shows that these vertices generate the annihilation and creation components of the projected dark-sector jump operators.

\subsection{Zeno elimination of the bright sector}
\label{sec:SM_zeno_elimination}

The microscopic Lindblad operators $L_\sigma=\sqrt{\kappa} c_\sigma $ empty the bright modes on the time scale $\kappa^{-1}$. 
In the Zeno regime, the dark modes therefore couple only through short-lived virtual excursions into the bright sector. 
We rewrite $H_{cf}$ in Eq.~\eqref{eq:SM_Hcf} as
\begin{equation}
H_{cf}
=
\sum_\sigma
\left(
 c_\sigma^\dagger\Lambda_\sigma
 +
 \Lambda_\sigma^\dagger c_\sigma
\right),
\label{eq:SM_Hcf_vertex}
\end{equation}
with $\Lambda_\sigma=-it e^{i\chi}\sin\chi f_\sigma\pm i\Gamma\sin\eta f^\dagger_{-\sigma}$, the upper sign holding for $\sigma=\uparrow$.  
This is the dark-sector vertex attached to the lossy bright mode $c_\sigma$.

We expand in the coupling of Eq.~\eqref{eq:SM_Hcf_vertex} to second order: one vertex moves a particle from the dark sector into the bright sector, the bright mode propagates, and a second vertex brings it back. 
The propagator connecting the two vertices must therefore be evaluated at zeroth order in $H_{cf}$, exactly as the auxiliary propagator of Sec.~\ref{sec:SM_auxiliary} was evaluated at $g=0$:
\begin{equation}
\mathsf{g}_{c,\sigma\sigma'}^R(\omega)
\equiv
-i\int_0^\infty e^{i\omega t}dt
\left\langle
\left\{c_\sigma(t),c_{\sigma'}^\dagger(0)\right\}
\right\rangle_c
=
-\frac{2i}{\kappa}\delta_{\sigma\sigma'}
+O(\kappa^{-2}),
\label{eq:SM_bright_Zeno_GF}
\end{equation}
for frequencies and coherent energy scales small compared with $\kappa$. 
Here $\langle\cdots\rangle_c$ denotes the stationary two-time average of the bright sector at $H_{cf}=0$, computed by the quantum regression theorem as in Eq.~\eqref{eq:SM_aux_gR_time}. 
The bright dispersion, pairing, and frequency dependence enter only at the next order. 
Integrating out the bright fields connects two vertices through Eq.~\eqref{eq:SM_bright_Zeno_GF}. 
In operator form, the induced retarded contribution to the quadratic dark-sector generator is
\begin{equation}
\Sigma_f^R(0)
=
\sum_{\sigma\sigma'}
\Lambda_\sigma^\dagger
\mathsf{g}_{c,\sigma\sigma'}^R(0)
\Lambda_{\sigma'}
=
-\frac{2i}{\kappa}
\sum_\sigma
\Lambda_\sigma^\dagger\Lambda_\sigma
+O(\kappa^{-2}).
\label{eq:SM_dark_selfenergy}
\end{equation}
To leading order, Eq.~\eqref{eq:SM_dark_selfenergy} is purely dissipative. 
A Hermitian correction to $H_f$ first appears at $O(\kappa^{-2})$.

The bright sector is empty to the order retained here. 
Strictly, the pairing term in $H_c$ produces a small stationary pair coherence and occupation even at $H_{cf}=0$. 
Their steady equations have the scaling
\begin{equation*}
0=-\kappa\langle c_\downarrow c_\uparrow\rangle_c+O(\Gamma),
\qquad
0=-\kappa\langle \hat n_{c\sigma}\rangle_c
+O\!\left(\Gamma\langle c_\downarrow c_\uparrow\rangle_c\right),
\end{equation*}
and hence $\langle c_\downarrow c_\uparrow\rangle_c=O(\Gamma/\kappa)$ and $\langle \hat n_{c\sigma}\rangle_c=O(\Gamma^2/\kappa^2)$. 
After integration over the bright correlation time $\kappa^{-1}$, the normal lesser kernel is $O(\Gamma^2/\kappa^3)$ and the anomalous contractions are at most $O(\Gamma/\kappa^2)$. 
Neither can contribute to the $O(\kappa^{-1})$ dark generator, so the lesser component may be set to zero at the present order. 
Equation~\eqref{eq:SM_dark_selfenergy} then has the same local loss structure derived in Sec.~\ref{sec:SM_auxiliary}: each decay vertex $\Lambda_\sigma$ gives, up to an irrelevant overall phase, a jump operator $\sqrt{4/\kappa}\,\Lambda_\sigma$. 
Equivalently,
\begin{equation}
\Sigma_f^R(0)
=
-\frac{i}{2}
\sum_\mu L_{f,\mu}^\dagger L_{f,\mu},
\label{eq:SM_selfenergy_lindblad_matching}
\end{equation}
and writing out the two vertices gives
\begin{equation}\label{eq:SM_Lf}
L_{f,1}
=
\sqrt{\frac{4}{\kappa}}
\left(
t\sin\chi f_\uparrow
-
\Gamma e^{-i\chi}\sin\eta f_\downarrow^\dagger
\right),
\qquad
L_{f,2}
=
\sqrt{\frac{4}{\kappa}}
\left(
t\sin\chi f_\downarrow
+
\Gamma e^{-i\chi}\sin\eta f_\uparrow^\dagger
\right).
\end{equation}
They generate the effective dark-sector master equation
\begin{equation}
\dot\rho_f
=
-i[H_f,\rho_f]
+
\mathcal{D}[L_{f,1}]\rho_f
+
\mathcal{D}[L_{f,2}]\rho_f
+
O(\kappa^{-2}),
\label{eq:SM_effective_dark_master}
\end{equation}
where $\rho_f=\operatorname{Tr}_c\rho$ to the order retained. 
The terms proportional to $f_\sigma$ describe dark loss. 
Those proportional to $f_\sigma^\dagger$ originate from the anomalous vertices in $H_{cf}$: a bright-dark pair is created and the bright fermion is removed by the reservoir. 
The next subsection resolves these effective Lindblad operators into quasiparticle loss and gain channels.

\subsection{Bogoliubov decomposition and secular limit}
\label{sec:SM_rates}

The occupation and the current require slightly different levels of approximation. 
The quasiparticle loss and gain rates determine the dark-level occupation even for the full projected dissipator, whereas the leading factorized current additionally uses a secular approximation to suppress quasiparticle pair coherences. 
To identify the rates and the regime in which that secular reduction is controlled, we diagonalize $H_f$ and re-express the jump operators in the resulting quasiparticle basis.

In the Nambu basis $\Psi_f=(f_\uparrow,f_\downarrow^\dagger)^T$, Eq.~\eqref{eq:SM_Hf} becomes
\begin{equation}
H_f
=
\begin{pmatrix}
f_\uparrow^\dagger & f_\downarrow
\end{pmatrix}
\begin{pmatrix}
-t\cos\chi & \Gamma e^{-i\chi}\cos\eta\\
\Gamma e^{i\chi}\cos\eta & t\cos\chi
\end{pmatrix}
\begin{pmatrix}
f_\uparrow \\
f_\downarrow^\dagger
\end{pmatrix}
+\text{const.},
\label{eq:SM_hf_matrix}
\end{equation}
with eigenvalues $\pm E_A$, where $E_A=\sqrt{t^2\cos^2\chi+\Gamma^2\cos^2\eta}=\Gamma\mathcal{E}_A$ and $\mathcal{E}_A\equiv\sqrt{r^2\cos^2\chi+\cos^2\eta}$. 
This is the Andreev bound state of the junction projected onto the dark sector (hence the subscript $A$ in $E_A$).  
We call it the dark Andreev level below. 
Diagonalizing Eq.~\eqref{eq:SM_hf_matrix} directly would give a (complex) Bogoliubov transformation. 
Rephasing the second Nambu component removes that phase and leaves a real symmetric matrix, which is diagonalized by an ordinary rotation. 
With $\mathcal{U}_\chi=\operatorname{diag}(1,e^{i\chi})$,
\begin{equation}
\mathcal{U}_\chi^\dagger
\begin{pmatrix}
-t\cos\chi & \Gamma e^{-i\chi}\cos\eta\\
\Gamma e^{i\chi}\cos\eta & t\cos\chi
\end{pmatrix}
\mathcal{U}_\chi
=
\begin{pmatrix}
-t\cos\chi & \Gamma\cos\eta\\
\Gamma\cos\eta & t\cos\chi
\end{pmatrix},
\end{equation}
so the remaining Bogoliubov rotation is real in the rephased basis $(f_\uparrow,\,e^{-i\chi}f^\dagger_\downarrow)^T$. 
Writing the result in terms of the original fields gives
\begin{equation}\label{eq:SM_Bogoliubov}
f_\uparrow
=
u\gamma_\uparrow
+e^{-i\chi}v\gamma_\downarrow^\dagger,
\qquad
f_\downarrow
=
u\gamma_\downarrow
-e^{-i\chi}v\gamma_\uparrow^\dagger,
\end{equation}
where $u,v\in\mathbb R$ obey
\begin{equation}
u^2+v^2=1,
\qquad
u^2-v^2=\frac{t\cos\chi}{E_A},
\qquad
2uv=\frac{\Gamma\cos\eta}{E_A}.
\label{eq:SM_uv}
\end{equation}
The Hamiltonian becomes
\begin{equation}
H_f
=
-E_A\sum_\sigma\gamma_\sigma^\dagger\gamma_\sigma
+\text{const.}
\label{eq:SM_Hf_diag}
\end{equation}
We use $\gamma_\sigma$ for the negative-energy branch throughout.

Substituting Eq.~\eqref{eq:SM_Bogoliubov} into the jump operators of Eq.~\eqref{eq:SM_Lf} gives
\begin{subequations}
\label{eq:SM_Lf_gamma}
\begin{align}
L_{f,1}
&=
\sqrt{\frac{4}{\kappa}}
\left(
A_-\gamma_\uparrow
-e^{-i\chi}A_+\gamma_\downarrow^\dagger
\right),
\\
L_{f,2}
&=
\sqrt{\frac{4}{\kappa}}
\left(
A_-\gamma_\downarrow
+e^{-i\chi}A_+\gamma_\uparrow^\dagger
\right),
\end{align}
\end{subequations}
with the real amplitudes
\begin{equation}
A_-
=
t\sin\chi u+\Gamma\sin\eta v,
\qquad
A_+
=
\Gamma\sin\eta u-t\sin\chi v.
\label{eq:SM_Apm}
\end{equation}
Because $\mathcal{D}[L]$ is quadratic in $L$, the cross terms between the annihilation and creation parts survive. 
Expanding the dissipators therefore produces three types of terms:
\begin{equation}
\sum_{\mu=1,2}\mathcal{D}[L_{f,\mu}]
=
\Gamma_-\sum_\sigma\mathcal{D}[\gamma_\sigma]
+
\Gamma_+\sum_\sigma\mathcal{D}[\gamma_\sigma^\dagger]
+
\mathcal{D}_{\rm pair},
\label{eq:SM_exact_dissipator_decomposition}
\end{equation}
with $\Gamma_-=4A_-^2/\kappa$ and $\Gamma_+=4A_+^2/\kappa$. 
The interference term is
\begin{equation}
\begin{aligned}
\mathcal{D}_{\rm pair}\rho
=
\frac{4A_-A_+}{\kappa}
\Big[
e^{i\chi}
\left(
\gamma_\downarrow\rho\gamma_\uparrow
-
\gamma_\uparrow\rho\gamma_\downarrow
\right)
+
e^{-i\chi}
\left(
\gamma_\uparrow^\dagger\rho\gamma_\downarrow^\dagger
-
\gamma_\downarrow^\dagger\rho\gamma_\uparrow^\dagger
\right)
+
\left\{
e^{-i\chi}\gamma_\uparrow^\dagger\gamma_\downarrow^\dagger
+
e^{i\chi}\gamma_\downarrow\gamma_\uparrow,
\rho
\right\}
\Big].
\end{aligned}
\label{eq:SM_pair_dissipator}
\end{equation}
The first two terms describe single-quasiparticle loss and gain at rates $\Gamma_-$ and $\Gamma_+$, respectively.
The third term, $\mathcal{D}_{\rm pair}$, couples density-matrix sectors whose quasiparticle numbers differ by two and generates pair coherences.
It arises from interference between the annihilation and creation parts of the same Lindblad operators, not from an independent two-particle loss or gain channel. 
It is also $O(\kappa^{-1})$, the same order as the terms we keep, so the Zeno expansion cannot remove it and a separate argument is needed. 
The explicit phases $e^{\pm i\chi}$ appear only in $\mathcal{D}_{\rm pair}$, while the rates depend on phase through $A_\pm$.

That argument is a separation of time scales.  
Under the diagonal Hamiltonian in Eq.~\eqref{eq:SM_Hf_diag},
\begin{equation}
\gamma_\sigma(t)=e^{iE_At}\gamma_\sigma,
\qquad
\gamma_\sigma^\dagger(t)=e^{-iE_At}\gamma_\sigma^\dagger.
\end{equation}
The loss and gain dissipators are stationary, whereas every term in $\mathcal{D}_{\rm pair}$ oscillates as $e^{\pm2iE_At}$. 
If the dissipative rates are slow compared with that oscillation, the pair term averages to zero on time scales that resolve the oscillation but leave the populations frozen. 
Writing $\Gamma_{\rm diss}\equiv\Gamma_-+\Gamma_+$, the condition $\Gamma_{\rm diss}\ll2E_A$ admits such a coarse-graining window, $(2E_A)^{-1}\ll\Delta t\ll\Gamma_{\rm diss}^{-1}$.
Averaging over it is the secular approximation, and it leaves
\begin{equation}
\sum_{\mu=1,2}\mathcal{D}[L_{f,\mu}]
\longrightarrow
\Gamma_-\sum_\sigma\mathcal{D}[\gamma_\sigma]
+
\Gamma_+\sum_\sigma\mathcal{D}[\gamma_\sigma^\dagger].
\label{eq:SM_secular_dissipator}
\end{equation}
The pair-coherence scale is $4|A_-A_+|/\kappa=\sqrt{\Gamma_-\Gamma_+}\leq\Gamma_{\rm diss}$, so the same condition controls all discarded terms. 
Since $\Gamma_{\rm diss}=O(\kappa^{-1})$, the secular limit is controlled at fixed $E_A>0$ as $\kappa\to\infty$, and the window closes near the isolated degeneracy $E_A=0$.  
What is lost there is the suppression of the pair coherences, not the rate equation for the level occupation. 
In the next subsection we first show that the occupation equation survives the full dissipator and then identify where the secular approximation enters the factorized current.

\subsection{Occupation imbalance and factorized current}
\label{sec:SM_current}

We first derive the occupation from the secular generator and then show that the resulting rate equation is in fact exact for the full projected dissipator. 
We subsequently evaluate the current, for which the secular suppression of pair coherences is required at leading order. 
Together with the diagonal Hamiltonian in Eq.~\eqref{eq:SM_Hf_diag}, Eq.~\eqref{eq:SM_secular_dissipator} gives
\begin{equation}
\dot\rho_f
=
-i\left[H_f,\rho_f\right]
+
\Gamma_-\sum_\sigma\mathcal{D}[\gamma_\sigma]\rho_f
+
\Gamma_+\sum_\sigma\mathcal{D}[\gamma_\sigma^\dagger]\rho_f
\equiv
\mathcal{L}_f\rho_f.
\label{eq:SM_dark_secular_master}
\end{equation}
The same amplitudes $A_\mp$ appear in both jump operators, so $\Gamma_\pm$ carry no spin label and the two spin components obey identical equations. 
Their stationary state is unique for $\Gamma_-+\Gamma_+>0$, so one symbol suffices,
\begin{equation}
n_\gamma
=
\left\langle
\gamma_\sigma^\dagger\gamma_\sigma
\right\rangle.
\label{eq:SM_ngamma_def}
\end{equation}
Expectation values evolve under the adjoint generator, constructed as in Sec.~\ref{sec:SM_auxiliary}: for any operator $O$,
\begin{equation}
\frac{d}{dt}\langle O\rangle
=
\operatorname{Tr}\left[O\dot\rho_f\right]
=
\operatorname{Tr}\left[O\mathcal{L}_f\rho_f\right]
=
\langle\mathcal{L}_f^\dagger(O) \rangle.
\end{equation}
Here $\mathcal{L}_f^\dagger$ is the adjoint of $\mathcal{L}_f$ with respect to the trace pairing, defined by $\operatorname{Tr}[O\mathcal{L}_f(\rho)] =
\operatorname{Tr}[\mathcal{L}_f^\dagger(O)\rho]$
for arbitrary $O$ and $\rho$. 
Because $H_f$ commutes with $\gamma_\sigma^\dagger\gamma_\sigma$, only the dissipative terms of $\mathcal{L}_f^\dagger$ act on it, and
\begin{equation}
\mathcal{L}_f^\dagger\!\left(\gamma_\sigma^\dagger\gamma_\sigma\right)
=
\underbrace{
\Gamma_-\left(
\gamma_\sigma^\dagger\gamma_\sigma^\dagger\gamma_\sigma\gamma_\sigma
-\tfrac{1}{2}\left\{\gamma_\sigma^\dagger\gamma_\sigma,\gamma_\sigma^\dagger\gamma_\sigma\right\}
\right)}_{=-\Gamma_-\gamma_\sigma^\dagger\gamma_\sigma}
+
\underbrace{
\Gamma_+\left(
\gamma_\sigma\gamma_\sigma^\dagger\gamma_\sigma\gamma_\sigma^\dagger
-\tfrac{1}{2}\left\{\gamma_\sigma\gamma_\sigma^\dagger,\gamma_\sigma^\dagger\gamma_\sigma\right\}
\right)}_{=\Gamma_+\left(1-\gamma_\sigma^\dagger\gamma_\sigma\right)},
\end{equation}
where $\gamma_\sigma\gamma_\sigma=0$ kills the loss jump term and $\gamma_\sigma\gamma_\sigma^\dagger=1-\gamma_\sigma^\dagger\gamma_\sigma$ is a projector.  
The factor $1-\gamma_\sigma^\dagger\gamma_\sigma$ in the gain channel is Pauli blocking. 
Taking the expectation value of this identity,
\begin{equation}
\dot n_\gamma
=
\Gamma_+(1-n_\gamma)
-
\Gamma_-n_\gamma,
\label{eq:SM_ngamma_rate}
\end{equation}
and the stationary occupation is
\begin{equation}
n_\gamma^{\rm ss}
=
\frac{\Gamma_+}{\Gamma_++\Gamma_-}.
\label{eq:SM_ngamma_ss}
\end{equation}
Equations~\eqref{eq:SM_ngamma_rate} and \eqref{eq:SM_ngamma_ss} do not in fact rely on the secular approximation. 
Repeating the calculation with the full dissipator of Eq.~\eqref{eq:SM_exact_dissipator_decomposition}, the pair contributions to $\mathcal{L}_f^\dagger(\gamma_\sigma^\dagger\gamma_\sigma)$ from $L_{f,1}$ and $L_{f,2}$ are equal and opposite and therefore cancel. 
The underlying reason is that Eq.~\eqref{eq:SM_Lf_gamma} is a pair of canonical fermion operators up to normalization,
\begin{equation*}
L_{f,1/2}
=
\sqrt{\Gamma_-+\Gamma_+}\;\tilde\gamma_{\uparrow/\downarrow},
\qquad
\tilde\gamma_\uparrow
=
\frac{A_-\gamma_\uparrow-e^{-i\chi}A_+\gamma_\downarrow^\dagger}{\sqrt{A_-^2+A_+^2}},
\qquad
\tilde\gamma_\downarrow
=
\frac{A_-\gamma_\downarrow+e^{-i\chi}A_+\gamma_\uparrow^\dagger}{\sqrt{A_-^2+A_+^2}},
\end{equation*}
with $\{\tilde\gamma_\sigma,\tilde\gamma_{\sigma'}^\dagger\}=\delta_{\sigma\sigma'}$ and $\{\tilde\gamma_\sigma,\tilde\gamma_{\sigma'}\}=0$. 
The exact dark dissipator is therefore pure loss of these rotated quasiparticles at the single rate $\Gamma_-+\Gamma_+$, and the level occupation is insensitive to $\mathcal{D}_{\rm pair}$ at any $E_A$. 
Both rates scale as $\kappa^{-1}$.  
The Zeno limit therefore increases the preparation time, $\tau_Z=(\Gamma_-+\Gamma_+)^{-1}\propto\kappa$, while leaving the gain-to-loss ratio $\Gamma_+/\Gamma_-$ independent of the overall $\kappa^{-1}$ scale whenever both rates are nonzero.

The balance between these gain and loss rates is encoded in the occupation imbalance
\begin{equation}
P_\gamma
\equiv
1-2n_\gamma^{\rm ss}
=
\frac{\Gamma_- - \Gamma_+}{\Gamma_-+\Gamma_+}
=
\frac{A_-^2-A_+^2}{A_-^2+A_+^2}.
\label{eq:SM_Pgamma}
\end{equation}
Thus $P_\gamma=+1$ for empty dark quasiparticle states, $P_\gamma=-1$ when both spin states are filled, and $P_\gamma=0$ when loss and gain balance. These occupations refer to the negative-energy Bogoliubov branch defined in Eq.~\eqref{eq:SM_Hf_diag}.
Unlike an equilibrium occupation, $P_\gamma$ is set by the reservoir-engineered rates and inherits their phase dependence. 
Using Eqs.~\eqref{eq:SM_uv} and \eqref{eq:SM_Apm}, we obtain
\begin{equation}
A_-^2-A_+^2
=
\frac{
 t\cos\chi
\left(t^2\sin^2\chi-\Gamma^2\sin^2\eta\right)
+
2t\Gamma^2\sin\chi\sin\eta\cos\eta
}{E_A},
\qquad
A_-^2+A_+^2
=
t^2\sin^2\chi+\Gamma^2\sin^2\eta.
\end{equation}
Equation~\eqref{eq:SM_Pgamma} then becomes
\begin{equation}
P_\gamma(\phi,\chi)
=
\frac{
r
\left[
\cos\chi\left(r^2\sin^2\chi-\sin^2\eta\right)
+2\sin\chi\sin\eta\cos\eta
\right]
}{
\mathcal{E}_A\left(r^2\sin^2\chi+\sin^2\eta\right)
}.
\label{eq:SM_Pgamma_closed}
\end{equation}
Equation~\eqref{eq:SM_ngamma_ss} requires only $\Gamma_-+\Gamma_+>0$, since it does not rest on the secular reduction. 
Equation~\eqref{eq:SM_Pgamma_closed} requires $E_A>0$ in addition, because $u$ and $v$ are expressed through $E_A$. 
If $A_-=A_+=0$, equivalently $t\sin\chi=\Gamma\sin\eta=0$, then $H_{cf}=0$ and the reservoir is exactly decoupled from the dark sector.  
For nonzero $t$ and $\Gamma$, this reduces to $\sin\chi=\sin\eta=0$.  
Its stationary occupation is therefore not uniquely selected.  
Because this decoupling is exact, finite $\kappa$ alone does not remove the ambiguity.

We now evaluate the current. 
The transport-current operator is $-2e\partial_\phi(H_c+H_f+H_{cf})$, where $H_c$ and $H_{cf}$ also depend on $\phi$ through $\cos\eta$ and $\sin\eta$.
Both contributions are down by one power of $\kappa$.  
Indeed, their steady equations have the schematic form
\begin{equation*}
0=-\kappa\langle c_\downarrow c_\uparrow\rangle+O(\Gamma),
\qquad
0=-\frac{\kappa}{2}\langle c_\sigma^\dagger f_{\sigma'}\rangle
+O(t,\Gamma),
\end{equation*}
with analogous equations for anomalous bright-dark coherences.
Thus every expectation value entering $\partial_\phi H_c$ or $\partial_\phi H_{cf}$ is $O(\kappa^{-1})$ at fixed $t$ and $\Gamma$, whereas the dark contribution is $O(\kappa^0)$.  
We therefore drop the former at the order retained here and keep only the dark level.  
In the symmetric gauge of Eq.~\eqref{eq:SM_symmetric_gauge},
\begin{equation}
I_f
=
-2e\frac{\partial H_f}{\partial\phi}
=
e\Gamma\sin\eta
\left(
e^{-i\chi}f_\uparrow^\dagger f_\downarrow^\dagger
+
e^{i\chi}f_\downarrow f_\uparrow
\right),
\label{eq:SM_current_operator_dark}
\end{equation}
where $\partial_\phi\eta=1/2$ has been used.  
For $\Gamma_-+\Gamma_+>0$, the secular generator has a unique steady state and is invariant under the global transformation $\gamma_\sigma\to e^{i\theta}\gamma_\sigma$.  
Consequently,
\begin{equation}
\langle\gamma_\downarrow\gamma_\uparrow\rangle_{\rm ss}
=
\langle\gamma_\uparrow^\dagger\gamma_\downarrow^\dagger\rangle_{\rm ss}
=0,
\qquad
\langle\gamma_\sigma^\dagger\gamma_\sigma\rangle_{\rm ss}
=n_\gamma^{\rm ss}.
\label{eq:SM_gamma_steady_correlators}
\end{equation}
This is where the secular approximation enters the current.  
The full dissipator of Eq.~\eqref{eq:SM_exact_dissipator_decomposition} is not invariant under $\gamma_\sigma\to e^{i\theta}\gamma_\sigma$, since $\mathcal{D}_{\rm pair}$ picks up $e^{\pm2i\theta}$, and it does generate a nonvanishing pair coherence.  
Unlike the occupation, which is exact, the anomalous correlators in Eq.~\eqref{eq:SM_gamma_steady_correlators} are set to zero only by the secular reduction.  
Away from $E_A=0$, keeping them would change the steady-state density matrix by a relative amount $\sqrt{\Gamma_-\Gamma_+}/(2E_A)$, which is $O(\kappa^{-1})$.  
The Bogoliubov transformation in Eq.~\eqref{eq:SM_Bogoliubov} therefore gives, to leading order,
\begin{equation}\label{eq:SM_anomalous_f_correlators}
\langle f_\downarrow f_\uparrow \rangle_{\rm ss}
=
e^{-i\chi}uv
\left(1-2n_\gamma^{\rm ss}\right)
=
e^{-i\chi}uvP_\gamma,
\qquad
\langle f_\uparrow^\dagger f_\downarrow^\dagger \rangle_{\rm ss}
=
e^{i\chi}uvP_\gamma.
\end{equation}
The steady current is the stationary expectation value $I_{\rm ss}=\langle I_f\rangle_{\rm ss}$.  
Substituting Eq.~\eqref{eq:SM_anomalous_f_correlators} into Eq.~\eqref{eq:SM_current_operator_dark} and using Eq.~\eqref{eq:SM_uv}, we obtain
\begin{equation}
I_{\rm ss}(\phi,\chi)
=
2e\Gamma\sin\eta uvP_\gamma
=
e\frac{\Gamma^2\sin\eta\cos\eta}{E_A}P_\gamma
=
e\Gamma\frac{\sin\eta\cos\eta}{\mathcal{E}_A}P_\gamma.
\label{eq:SM_Iss_uv}
\end{equation}
Thus, to leading order in the Zeno regime under the secular approximation, the steady current factorizes as
\begin{equation}
I_{\rm ss}(\phi,\chi)
=
I_A(\phi,\chi)P_\gamma(\phi,\chi),
\qquad
I_A(\phi,\chi)
=
e\frac{\Gamma^2\sin\eta\cos\eta}{E_A}
=
e\Gamma\frac{\sin\eta\cos\eta}{\mathcal{E}_A}.
\label{eq:SM_factorized_current}
\end{equation}
Combining Eqs.~\eqref{eq:SM_factorized_current} and \eqref{eq:SM_Pgamma_closed} gives the closed current-phase relation (CPR)
\begin{equation}
I_{\rm ss}(\phi,\chi)
=
e\Gamma
\frac{
r\sin\eta\cos\eta
\left[
\cos\chi\left(r^2\sin^2\chi-\sin^2\eta\right)
+
2\sin\chi\sin\eta\cos\eta
\right]
}{
\left(r^2\cos^2\chi+\cos^2\eta\right)
\left(r^2\sin^2\chi+\sin^2\eta\right)
}.
\label{eq:SM_Iss_closed}
\end{equation}
Setting $r=1$, with energies measured in units of $\Gamma$ and currents in units of $e\Gamma$, gives the normalization used in the main text.
The factorization isolates the origin of the diode response. 
The spectral factor $I_A$ is the current carried by the surviving dark Andreev level.  
It is odd under $\eta\to-\eta$ while $E_A$ is even, so $\max_\phi I_A=-\min_\phi I_A$: taken alone it is reciprocal, with equal positive and negative critical magnitudes. 
The reservoir-selected occupation imbalance $P_\gamma$ reweights this reciprocal spectral current asymmetrically in phase.
Within the noninteracting infinite-gap Zeno and secular approximation, rectification therefore originates entirely from occupation reweighting.

\subsection{Robustness to pairing asymmetry and hopping variation}
\label{sec:SM_asymmetry_robustness}

The closed-form result above holds at $\Gamma_L=\Gamma_R$ for arbitrary $r=t/\Gamma$. 
We now solve the full infinite-gap Lindblad equation at finite $\kappa$ to test two complementary variations: breaking the pairing symmetry through $\Gamma_R/\Gamma_L$, and changing $t/\Gamma$ while keeping $\Gamma_L=\Gamma_R$. 
For each parameter set, we define the forward and backward critical currents
\begin{equation}
I_c^+
=
\max\!\left[0,\max_\phi I_{\rm ss}(\phi)\right],
\qquad
|I_c^-|
=
\max\!\left[0,-\min_\phi I_{\rm ss}(\phi)\right],
\label{eq:SM_Ic_asymmetry}
\end{equation}
and the diode efficiency
\begin{equation}
\eta_D
=
\frac{I_c^+-|I_c^-|}{I_c^++|I_c^-|}.
\label{eq:SM_etaD_asymmetry}
\end{equation}

\begin{figure}
\centering
\includegraphics[width=0.98\linewidth]{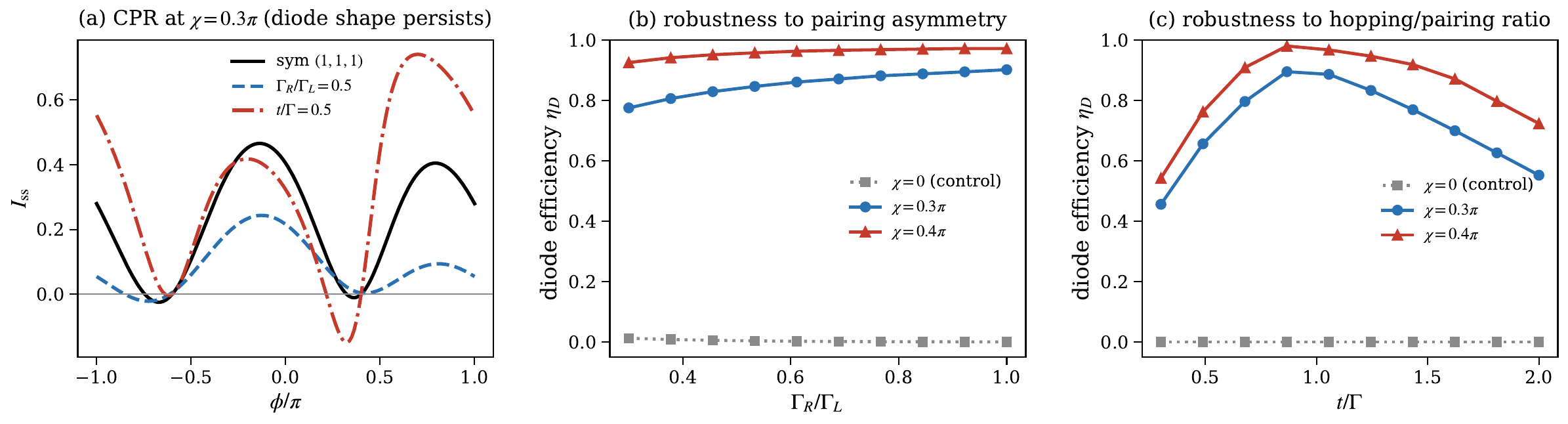}
\caption{
Robustness to pairing asymmetry and hopping variation. 
All data are obtained from the exact infinite-gap Lindblad steady state at $\kappa=300$. 
(a) Steady current-phase relation $I_{\rm ss}(\phi)$ at $\chi=0.3\pi$ for the reference point $\Gamma_R/\Gamma_L=t/\Gamma=1$, pairing asymmetry $\Gamma_R/\Gamma_L=0.5$, and hopping variation $t/\Gamma=0.5$.
(b) Diode efficiency $\eta_D$ versus $\Gamma_R/\Gamma_L$. 
(c) Diode efficiency $\eta_D$ versus $t/\Gamma$. 
The $\chi=0$ control remains nearly reciprocal, while finite reservoir phases retain a large diode efficiency over broad parameter ranges.
}
\label{fig:SM_asymmetry_robustness}
\end{figure}

Figure~\ref{fig:SM_asymmetry_robustness} shows the result for $\kappa=300$. 
Panel (a) compares the steady CPR at $\chi=0.3\pi$ for the symmetric reference point $\Gamma_R/\Gamma_L=1$ and $t/\Gamma=1$, for pairing asymmetry $\Gamma_R/\Gamma_L=0.5$, and for the hopping variation $t/\Gamma=0.5$. 
Although the shape and magnitude of the CPR change quantitatively, the diode form survives: the backward branch remains strongly depleted while the forward branch stays finite. 
In panel (b), we fix $t=\Gamma_L=1$ and vary $\Gamma_R$, thereby breaking the pairing symmetry.  
In panel (c), we set $\Gamma_L=\Gamma_R\equiv\Gamma=1$ and vary $t$, directly illustrating the finite-$\kappa$ response across the ratio $r=t/\Gamma$ already retained in the analytic theory.  
The diode response survives pairing asymmetry and remains strong over a broad hopping range: the $\chi=0$ control stays nearly reciprocal, whereas finite reservoir phases retain a large $\eta_D$.

\subsection{Interaction-induced dark-pair resonance and two-particle loss}
\label{sec:interaction_benchmark}

Returning to $\Gamma_L=\Gamma_R$, we now restore the local Hubbard interaction and work in the Zeno regime $\kappa \gg |t|,\Gamma, U$. 
We first isolate the purely dark component of the interaction, which acts at $O(\kappa^0)$, and then integrate out the bright-pair channel with the same bare-propagator expansion used for a single bright fermion in Sec.~\ref{sec:SM_zeno_elimination}. 
The resulting interacting dark-sector master equation is finally benchmarked against the full microscopic dynamics.

We start with the microscopic interaction in the original basis,
\begin{equation}
H_U
=
U\sum_{j=L,R}\hat n_{j\uparrow}\hat n_{j\downarrow},
\qquad
\hat n_{j\sigma}=d_{j\sigma}^\dagger d_{j\sigma}.
\label{eq:HU_appendix}
\end{equation}
Substituting Eq.~\eqref{eq:SM_inverse_bright_dark} into Eq.~\eqref{eq:HU_appendix} and collecting like terms yields
\begin{equation}
H_U
=
H_{U,f}+H_{U,{\rm pair}}+H_{U,{\rm rem}},
\label{eq:HU_exact_split}
\end{equation}
where
\begin{subequations}
\begin{align}
H_{U,f}
&=
\frac{U}{2}\hat n_{f\uparrow}\hat n_{f\downarrow},
\label{eq:HU_dark_final}
\\
H_{U,{\rm pair}}
&=
\frac{U}{2}
\left(
 e^{2i\chi}\Pi_c^\dagger\Pi_f
+
 e^{-2i\chi}\Pi_f^\dagger\Pi_c
\right),
\qquad
\Pi_c=c_\downarrow c_\uparrow,
\quad
\Pi_f=f_\downarrow f_\uparrow,
\label{eq:interaction_pair_vertex}
\\
H_{U,{\rm rem}}
&=
\frac{U}{2}
\Bigl[
 \hat n_{c\uparrow}\hat n_{c\downarrow}
+\hat n_{c\uparrow}\hat n_{f\downarrow}
+\hat n_{f\uparrow}\hat n_{c\downarrow}
-c_\uparrow^\dagger c_\downarrow
 f_\downarrow^\dagger f_\uparrow
-c_\downarrow^\dagger c_\uparrow
 f_\uparrow^\dagger f_\downarrow
\Bigr].
\label{eq:interaction_remainder}
\end{align}
\end{subequations}
Of the three terms, the first acts only on the dark modes, the second converts a dark pair into a bright pair or the reverse, and the third contains a bright occupation or a bright spin-flip operator, so $H_{U,{\rm rem}}$ vanishes when the bright modes are empty.
We return below to the order at which this remainder can affect the projected dynamics.

The first interaction-induced effect is the direct coherent action of the purely dark term $H_{U,f}$, which is present at $O(\kappa^0)$. 
Adding $H_{U,f}$ of Eq.~\eqref{eq:HU_dark_final} to $H_f$ of Eq.~\eqref{eq:SM_Hf} gives the zeroth-order dark-sector Hamiltonian in the $1/\kappa$ expansion:
\begin{equation}
H_{U,f}^{(0)}
=
H_f+H_{U,f}
=
-t\cos\chi\sum_\sigma \hat n_{f\sigma}
+
\Gamma\cos\eta
\left(
 e^{-i\chi}f_\uparrow^\dagger f_\downarrow^\dagger
+
 e^{i\chi}f_\downarrow f_\uparrow
\right)
+
\frac{U}{2}\hat n_{f\uparrow}\hat n_{f\downarrow}.
\label{eq:interacting_Hf_general}
\end{equation}
Fermion parity block-diagonalizes the above Hamiltonian. 
In the even-parity basis $\{|0\rangle_f,|\!\uparrow\downarrow\rangle_f\}$, the nontrivial block is
\begin{equation}
H_{\rm even}^{(0)}
=
\begin{pmatrix}
0 & \Gamma_\eta e^{i\chi}\\
\Gamma_\eta e^{-i\chi} & \delta_U
\end{pmatrix},
\qquad
\delta_U=-2t\cos\chi+\frac{U}{2},
\qquad
\Gamma_\eta\equiv\Gamma\cos\eta.
\label{eq:Heven_appendix_final}
\end{equation}
In the odd-parity basis $\{| \uparrow \rangle_f,| \downarrow \rangle_f\}$, both the pairing term and the projected interaction vanish, leaving $ H_{\rm odd}^{(0)} = -t\cos\chi\,\mathbb{I}_2$. 
This block is proportional to the identity and therefore does not participate in the resonance between the empty and doubly occupied states.
From Eq.~\eqref{eq:Heven_appendix_final}, the empty and doubly occupied states are resonant at
\begin{equation}
\delta_U=0
\quad\Longrightarrow\quad
U_{\rm res}=4t\cos\chi.
\label{eq:Ures_appendix_final}
\end{equation}
For repulsive $U$, this resonance lies at positive interaction when $\cos\chi>0$. 
A finite $\Gamma_\eta$ turns the crossing into an avoided crossing. 
This resonance follows from the purely dark component of the interaction and does not involve a virtual bright state or a bright-sector propagator.

We now turn to the second interaction-induced effect, which arises at $O(\kappa^{-1})$ from a virtual excursion through the lossy bright-pair sector. 
Two Hamiltonian terms generate this excursion: the proximity-induced pair term $H_{c,{\rm pair}}$ from the pairing part of Eq.~\eqref{eq:SM_Hc}, and the interaction-induced pair-conversion term $H_{U,{\rm pair}}$ in Eq.~\eqref{eq:interaction_pair_vertex}.  
Collecting them, we write
\begin{equation}
\begin{aligned}
V_{\rm bp}
\equiv
H_{c,{\rm pair}}+H_{U,{\rm pair}}
=
\Pi_c^\dagger\Lambda_{\rm bp}
+
\Lambda_{\rm bp}^\dagger\Pi_c,
\qquad
\text{with}
\qquad
\Lambda_{\rm bp}
=
\Gamma_\eta e^{i\chi}
+
\frac{U}{2}e^{2i\chi}\Pi_f.
\end{aligned}
\label{eq:interaction_bright_pair_vertex}
\end{equation}
$V_{\rm bp}$ is the perturbation that creates and annihilates the virtual bright pair. 
The bare propagator of the bright pair is obtained by setting $V_{\rm bp}=0$ while retaining the bright number term and the one-body loss. 
Denoting this bare bright-sector generator by $\mathcal{L}_{c,0}$, its adjoint acts on the pair annihilation operator as
\begin{equation*}
\mathcal{L}_{c,0}^\dagger(\Pi_c)
=
-\left(2it\cos\chi+\kappa\right)\Pi_c.
\end{equation*}
The frequency $2t\cos\chi$ is the energy of two bright fermions, while the pair amplitude decays at $\kappa$ because each constituent amplitude decays at $\kappa/2$. 
In direct analogy with the single-bright-fermion propagator $\mathsf g_c^R$ in Sec.~\ref{sec:SM_zeno_elimination}, we denote the bare retarded propagator of this composite channel by $\mathsf{g}_{\rm bp}^R$, where ``bp'' stands for bright pair:
\begin{equation*}
\mathsf{g}_{\rm bp}^R(\tau)
\equiv
-i\theta(\tau)
\left\langle
\left[\Pi_c(\tau),\Pi_c^\dagger(0)\right]
\right\rangle_{c,0}.
\end{equation*}
The commutator is used because $\Pi_c$ has even fermion parity. 
In the bare bright vacuum, $\langle\Pi_c^\dagger\Pi_c\rangle_{c,0}=0$ and $\langle\Pi_c\Pi_c^\dagger\rangle_{c,0}=1$. 
The quantum regression theorem then gives $\mathsf{g}_{\rm bp}^R(\tau)=-i\theta(\tau)e^{-(2it\cos\chi+\kappa)\tau}$, whose Fourier transform is
\begin{equation}
\mathsf{g}_{\rm bp}^R(\omega)
=
\frac{1}{\omega-2t\cos\chi+i\kappa},
\qquad
\mathsf{g}_{\rm bp}^R(0)
=
\frac{1}{-2t\cos\chi+i\kappa}
=
-\frac{i}{\kappa}
-
\frac{2t\cos\chi}{\kappa^2}
+O(\kappa^{-3}).
\label{eq:bright_pair_GF}
\end{equation}
At second order in $V_{\rm bp}$, $\Pi_c^\dagger\Lambda_{\rm bp}$ creates the virtual bright pair, the pair propagates according to Eq.~\eqref{eq:bright_pair_GF}, and $\Lambda_{\rm bp}^\dagger\Pi_c$ annihilates it.  
This process gives the retarded self-energy
\begin{equation}
\begin{aligned}
\Sigma_{\rm bp}^R(\omega)
&=
\Lambda_{\rm bp}^\dagger
\mathsf{g}_{\rm bp}^R(\omega)
\Lambda_{\rm bp},
\\
\Sigma_{\rm bp}^R(0)
&=
-\frac{i}{\kappa}\Lambda_{\rm bp}^\dagger\Lambda_{\rm bp}
+O(\kappa^{-2}).
\end{aligned}
\label{eq:interaction_pair_selfenergy}
\end{equation}
The bare bright-pair sector is empty, so its lesser contraction vanishes.  
Matching Eq.~\eqref{eq:interaction_pair_selfenergy} to a Lindblad channel in the same way as Eq.~\eqref{eq:SM_selfenergy_lindblad_matching} immediately gives the jump operator $L_{\rm bp}=\sqrt{2/\kappa}\,\Lambda_{\rm bp}$.
\begin{samepage}
Writing
\begin{equation}
L_{\rm bp}
=
\alpha_U+\beta_U\Pi_f,
\qquad
\alpha_U
=
\sqrt{\frac{2}{\kappa}}\Gamma_\eta e^{i\chi},
\qquad
\beta_U
=
\sqrt{\frac{2}{\kappa}}\frac{U}{2}e^{2i\chi},
\label{eq:interaction_pair_parts}
\end{equation}
the resulting dissipator first expands into
\begin{align}
\mathcal{D}[L_{\rm bp}]\rho_f
&=
|\alpha_U|^2\rho_f
+\alpha_U^*\beta_U\Pi_f\rho_f
+\alpha_U\beta_U^*\rho_f\Pi_f^\dagger
+|\beta_U|^2\Pi_f\rho_f\Pi_f^\dagger
\nonumber\\
&\quad
-
\frac{1}{2}
\Big\{
|\alpha_U|^2\mathbb{I}
+\alpha_U^*\beta_U\Pi_f
+\alpha_U\beta_U^*\Pi_f^\dagger
+|\beta_U|^2\Pi_f^\dagger\Pi_f,
\rho_f
\Big\}
\nonumber\\
&=
\underbrace{
|\alpha_U|^2\rho_f
-\frac{1}{2}\big\{|\alpha_U|^2\mathbb{I},\rho_f\big\}
}_{=0}
+|\beta_U|^2\mathcal{D}[\Pi_f]\rho_f
+
\frac{1}{2}
\left[
\alpha_U^*\beta_U\Pi_f
-\alpha_U\beta_U^*\Pi_f^\dagger,
\rho_f
\right]
\nonumber\\
&=
\mathcal{D}[L_{U,f}]\rho_f
-i[H_{U,f}^{(1)},\rho_f].
\label{eq:interaction_pair_decomposition}
\end{align}
The underbraced term vanishes because $\alpha_U$ multiplies the identity. 
The term $|\beta_U|^2\mathcal{D}[\Pi_f]$ is the dark-pair loss channel, while the cross term is a coherent correction to the dark Hamiltonian. 
Explicitly,
\begin{subequations}
\begin{align}
L_{U,f}
&=
\sqrt{\gamma_U}\Pi_f,
\qquad
\gamma_U
=
\frac{U^2}{2\kappa},
\label{eq:JU_appendix_final}
\\
H_{U,f}^{(1)}
&=
\frac{iU\Gamma_\eta}{2\kappa}
\left(
 e^{i\chi}\Pi_f
-
 e^{-i\chi}\Pi_f^\dagger
\right).
\label{eq:interaction_coherent_cross}
\end{align}
\end{subequations}
\end{samepage}
Both terms arise from the same eliminated bright-pair channel. 
The first removes a dark pair at rate $\gamma_U=O(U^2/\kappa)$, and the second changes the coherent pair amplitude at $O(U\Gamma/\kappa)$. 
The coherent correction $H_{U,f}^{(1)}$ contains only $\Pi_f$ and $\Pi_f^\dagger$ and therefore also acts within the even sector.  
With $
q_U=\frac{U\Gamma_\eta}{2\kappa},
$
the even-parity Hamiltonian is
\begin{equation}
H_{\rm even}
=
\begin{pmatrix}
0 & e^{i\chi}(\Gamma_\eta+iq_U)\\
e^{-i\chi}(\Gamma_\eta-iq_U) & \delta_U
\end{pmatrix}.
\label{eq:interaction_Heven_corrected}
\end{equation}
Thus $q_U$ changes the complex pairing matrix element but not the detuning $\delta_U$, so the resonance stays at $\delta_U=0$.  
The odd Hamiltonian is unchanged, $H_{\rm odd}^{(0)}=-t\cos\chi\,\mathbb{I}_2$. 
The odd states are not discarded from the open-system dynamics: the one-body operators $L_{f,1}$ and $L_{f,2}$ transfer population between the even and odd sectors.  
Their population is included below through $1-P_{\rm e}$, whereas only the even sector supports the pair coherence that carries the current.

The bright-pair elimination has therefore produced both the nonlinear loss $L_{U,f}$ and the coherent correction $H_{U,f}^{(1)}$ at the same order in $1/\kappa$. 
It remains to check whether any other interaction vertices modify the leading projected dynamics. 
The vertices in Eq.~\eqref{eq:SM_Hcf_vertex} contain one bright operator, whereas $H_{U,{\rm pair}}$ contains two.  
A second-order process with one vertex of each type cannot start and end in the bright vacuum, so the mixed kernel vanishes. 
Every term in $H_{U,{\rm rem}}$ also vanishes on the bright vacuum and can act only inside a virtual bright excursion.  
Such an insertion adds one bright propagator and starts at $O(\kappa^{-2})$. 
The interaction therefore leaves the one-body Lindblad operators unchanged at $O(\kappa^{-1})$: they remain $L_{f,1}$ and $L_{f,2}$ of Eq.~\eqref{eq:SM_Lf}, and no additional one-body channel appears at this order. 
Collecting the terms gives
\begin{equation}
\dot\rho_f
=
-i[H_{U,f}^{(0)}+H_{U,f}^{(1)},\rho_f]
+
\sum_{\mu=1,2}\mathcal{D}[L_{f,\mu}]\rho_f
+
\mathcal{D}[L_{U,f}]\rho_f
+
O(\kappa^{-2}).
\label{eq:interacting_dark_master}
\end{equation}
Equation~\eqref{eq:interacting_dark_master} is the central result of this subsection: the interacting dark-sector dynamics through $O(\kappa^{-1})$.

We next evaluate the supercurrent, which is fixed by the dark-pair coherence. 
Because $H_U$ has no explicit $\phi$ dependence, the leading dark current operator remains that of Eq.~\eqref{eq:SM_current_operator_dark}, whose expectation value depends only on
\begin{equation}
Z_{\rm ss}=X_{\rm ss}+iY_{\rm ss}
\equiv
e^{i\chi}\langle f_\downarrow f_\uparrow\rangle_{\rm ss}.
\label{eq:interacting_pair_coherence}
\end{equation}
For compactness, define
\begin{equation}
\lambda_\kappa=\frac{4}{\kappa},
\qquad
a=t\sin\chi,
\qquad
b=\Gamma\sin\eta,
\qquad
S=a^2+b^2,
\qquad
C=ab,
\qquad
A=\lambda_\kappa S+\frac{\gamma_U}{2}.
\label{eq:interacting_aux_defs}
\end{equation}
To close the equation for $Z_{\rm ss}$, we also use the empty and doubly occupied probabilities and their combinations
\begin{equation}
\begin{gathered}
p_0
=
\left\langle
(1-\hat n_{f\uparrow})(1-\hat n_{f\downarrow})
\right\rangle_{\rm ss},
\qquad
p_2
=
\langle\hat n_{f\uparrow}\hat n_{f\downarrow}\rangle_{\rm ss},
\\
P_{\rm e}=p_0+p_2,
\qquad
W=p_2-p_0.
\end{gathered}
\label{eq:interacting_observables}
\end{equation}
The total odd-parity population is $1-P_{\rm e}$.

Applying the adjoint of Eq.~\eqref{eq:interacting_dark_master} to $\Pi_f$ and to the population projectors in Eq.~\eqref{eq:interacting_observables} gives a closed steady-state system.  
Spin symmetry leaves only $X_{\rm ss}$, $Y_{\rm ss}$, $P_{\rm e}$, and $W$ independent.  
Before eliminating $P_{\rm e}$, the four real steady-state equations are
\begin{subequations}\label{eq:interacting_XYPeW_system}
\begin{align}
0
&=
-A X_{\rm ss}
+\delta_U Y_{\rm ss}
+q_U W
+\lambda_\kappa C,
\\
0
&=
-\delta_U X_{\rm ss}
-A Y_{\rm ss}
+\Gamma_\eta W,
\\
0
&=
\lambda_\kappa S(1-2P_{\rm e})
+\lambda_\kappa(b^2-a^2)W
+4\lambda_\kappa C X_{\rm ss},
\\
0
&=
-4q_U X_{\rm ss}
-4\Gamma_\eta Y_{\rm ss}
-\gamma_U P_{\rm e}
-(\lambda_\kappa S+\gamma_U)W
+\lambda_\kappa(b^2-a^2).
\end{align}
\end{subequations}
The first two equations are the real and imaginary parts of the pair-coherence equation, while the last two govern the total even-parity population and the population imbalance between the doubly occupied and empty states.
For $S>0$, the third equation gives
\begin{equation}
P_{\rm e}
=
\frac{1}{2}
+\frac{b^2-a^2}{2S}W
+\frac{2C}{S}X_{\rm ss}.
\label{eq:interacting_Pe_elimination}
\end{equation}
Substituting this result into the fourth equation, dividing it by two, and defining
\begin{equation}
B
=
\frac{\lambda_\kappa S}{2}
+
\frac{\gamma_U(a^2+3b^2)}{4S},
\qquad
K
=
\frac{\lambda_\kappa}{2}(b^2-a^2)
-
\frac{\gamma_U}{4},
\label{eq:interacting_BK_defs}
\end{equation}
gives
\begin{equation}
\begin{pmatrix}
A & -\delta_U & -q_U\\
\delta_U & A & -\Gamma_\eta\\
\dfrac{\gamma_UC}{S}+2q_U & 2\Gamma_\eta & B
\end{pmatrix}
\begin{pmatrix}
X_{\rm ss}\\Y_{\rm ss}\\W
\end{pmatrix}
=
\begin{pmatrix}
\lambda_\kappa C\\0\\K
\end{pmatrix}.
\label{eq:interacting_XY_linear}
\end{equation}
Solving only for the component needed by the current yields
\begin{equation}
X_{\rm ss}
=
\frac{
\lambda_\kappa C\left(AB+2\Gamma_\eta^2\right)
+
K\left(Aq_U+\Gamma_\eta\delta_U\right)
}{
B\left(A^2+\delta_U^2\right)
+
2A\left(\Gamma_\eta^2+q_U^2\right)
+
\dfrac{\gamma_UC}{S}
\left(Aq_U+\Gamma_\eta\delta_U\right)
}.
\label{eq:Xss_appendix_final}
\end{equation}
Substitution into Eq.~\eqref{eq:SM_current_operator_dark} gives the analytic projected current
\begin{equation}
I_{\rm ss}^{({\rm an})}(\phi,\chi,U)
=
2e\Gamma\sin\eta X_{\rm ss}
+
O(\kappa^{-1}).
\label{eq:Iss_interacting_analytic}
\end{equation}
Equations~\eqref{eq:Xss_appendix_final} and \eqref{eq:Iss_interacting_analytic} assume $S>0$.  
If $S=0$, then $L_{f,1}=L_{f,2}=0$ and the odd-parity populations are conserved, so the stationary state is not unique.  
The detuning $\delta_U$ fixes the resonance center, while $\lambda_\kappa S$, $\gamma_U$, $\Gamma_\eta$, and $q_U$ determine its width and line shape.  
The $O(\kappa^{-1})$ remainder in Eq.~\eqref{eq:Iss_interacting_analytic} contains subleading current-operator contributions and the steady-state correction generated by the omitted $O(\kappa^{-2})$ terms in the effective Liouvillian.

At exact dark-pair resonance, $\delta_U=0$, the current simplifies considerably.
For $\Gamma_\eta\ne0$, the leading terms in the numerator and denominator of Eq.~\eqref{eq:Xss_appendix_final} are $2\lambda_\kappa C\Gamma_\eta^2$ and $2A\Gamma_\eta^2$, respectively.
The leading Zeno pair coherence is therefore
\begin{equation}
X_{\rm ss}^{\rm res}
\simeq
\frac{\lambda_\kappa C}{A}
=
\frac{t\Gamma\sin\chi\sin\eta}
{t^2\sin^2\chi+\Gamma^2\sin^2\eta+U^2/16}.
\label{eq:Xss_interacting_resonance}
\end{equation}
This reduction also holds at $\Gamma_\eta=0$: then $q_U=0$, and the first equation of Eq.~\eqref{eq:interacting_XYPeW_system} gives $AX_{\rm ss}=\lambda_\kappa C$ directly.
The term $U^2/16=\gamma_U/(2\lambda_\kappa)$ comes from the induced pair loss and survives in the steady state because both dissipative rates scale as $1/\kappa$.
Using $U=U_{\rm res}=4t\cos\chi$ gives $t^2\sin^2\chi+U^2/16=t^2$.
Substitution into Eq.~\eqref{eq:Iss_interacting_analytic} then yields
\begin{equation}
I_{\rm ss}^{\rm res}(\phi,\chi)
=
\frac{2et\Gamma^2\sin\chi\,\sin^2\eta}
{t^2+\Gamma^2\sin^2\eta}
+O(\kappa^{-1}),
\label{eq:Iss_interacting_resonance}
\end{equation}
which is Eq.~\eqref{eq:resonant_current_main} of the main text.
For $\sin\chi\ne0$, the leading current has a fixed sign wherever it is nonzero, giving ideal rectification at resonance.

We finally benchmark the projected current against the microscopic interacting Lindblad equation, solved in the full $2^4=16$-dimensional Fock space,
\begin{equation}
0
=
-i[H+H_U,\rho_{\rm ss}]
+
\sum_\sigma \mathcal{D}[L_\sigma]\rho_{\rm ss}.
\label{eq:exact_interacting_lindblad}
\end{equation}
The microscopic equation contains only one-body loss. 
The coherent term $H_{U,f}^{(1)}$ and the two-particle Lindblad operator $L_{U,f}$ emerge after the bright pair is integrated out.

The comparison with the full microscopic dynamics is shown in Fig.~\ref{fig:interaction_resonance}(b) of the main text.

\section{Beyond the Zeno and infinite-gap limits}
\label{sec:SM_keldysh}

We now set $U=0$ and relax the two limits used above: finite $\kappa$ requires retaining the bright sector, and a finite superconducting gap gives frequency-dependent lead self-energies.
The full dot-lead Hamiltonian is quadratic and the Lindblad operators are linear, so the Green's-function equations of motion close at the one-particle level.
The Keldysh framework provides a common description of coherent and dissipative dynamics~\cite{Sieberer2016Keldysh}.
In the following, Sections~\ref{sec:SM_KL_contour} and \ref{sec:SM_KL_quadratic} map a linear Lindblad channel onto Keldysh self-energies, following Stefanucci's formulation~\cite{Stefanucci2024KadanoffBaym}, reproduced here only to keep the presentation self-contained.
Section~\ref{sec:SM_KL_nambu} assembles the finite-gap steady state in Nambu space, combining the equilibrium lead self-energies with the nonlocal loss self-energy in a Dyson equation. 
Section~\ref{sec:SM_KL_currents} derives the lead and loss currents and their continuity relation, splitting the lead currents into transport and common-mode components. 
Finally, Sec.~\ref{sec:SM_operating_window} returns to the infinite-gap model to isolate the finite-$\kappa$ crossover from the finite-gap corrections.

\subsection{Contour representation of the Lindblad evolution}
\label{sec:SM_KL_contour}

This subsection maps the Lindblad dynamics of the dot fields onto a contour-ordered evolution, from which the Lindblad self-energies are read off in Sec.~\ref{sec:SM_KL_quadratic}. 
The density matrix obeys the Lindblad equation
\begin{equation}
\dot\rho
=
-i[H,\rho]
+
\sum_\mu
\left(
L_\mu\rho L_\mu^\dagger
-
\frac{1}{2}
\{L_\mu^\dagger L_\mu,\rho\}
\right).
\label{eq:SM_KL_master_equation}
\end{equation}
In the present model $\mu=\sigma$ runs over the two spin projections, with $L_\sigma$ given by Eq.~\eqref{eq:SM_nonlocal_jump}, and the Lindblad operators act only on the dot fields.  
The superconducting leads are prepared in equilibrium at the initial contour time and evolve under quadratic Hamiltonians.  
Their degrees of freedom are integrated out exactly in Sec.~\ref{sec:SM_KL_nambu}.

The construction proceeds in two steps.  
We first recast the master equation in terms of the non-Hermitian Hamiltonian
\begin{equation}
H_{\rm NH}
=
H
-
\frac{i}{2}
\sum_\mu L_\mu^\dagger L_\mu,
\label{eq:SM_KL_nonhermitian_H}
\end{equation}
so that Eq.~\eqref{eq:SM_KL_master_equation} becomes
\begin{equation}
\dot\rho
=
-i
\left(
H_{\rm NH}\rho
-
\rho H_{\rm NH}^\dagger
\right)
+
\sum_\mu
L_\mu\rho L_\mu^\dagger.
\label{eq:SM_KL_master_nonhermitian}
\end{equation}
The first term acts separately on the two sides of $\rho$ and determines the coherent evolution and damping. 
The jump term connects them and determines how the damped states are populated.

As the second step, we unfold the density-matrix dynamics onto the closed time contour $C=C_-\cup C_+$ to represent these left and right actions within a single ordered evolution. 
The forward branch $C_-$ propagates the ket, the backward branch $C_+$ propagates the bra, and vertices connecting the two branches encode the jump term. 
This construction retains the spectral information from the coherent evolution and damping together with the distribution information from the jump processes, in a form suitable for a Dyson equation. 
Let $s(z)$ denote the branch-orientation sign and $\theta_\pm(z)$ the branch indicator functions:
\begin{equation}
s(z)=
\begin{cases}
+1,& z\in C_-,\\
-1,& z\in C_+,
\end{cases}
\qquad
\theta_-(z)=
\begin{cases}
1,& z\in C_-,\\
0,& z\in C_+,
\end{cases}
\qquad
\theta_+(z)=1-\theta_-(z),
\label{eq:SM_KL_contour_branches}
\end{equation}
and let $z^*$ denote the point on the opposite branch at the same real time. 
The generator of the resulting contour-ordered Lindblad evolution is
\begin{equation}
H_C(z,z^*)
=
H(z)
-
\frac{i}{2}s(z)\sum_\mu L_\mu^\dagger(z)L_\mu(z)
+
i\theta_-(z)\sum_\mu L_\mu^\dagger(z^*)L_\mu(z).
\label{eq:SM_KL_contour_generator}
\end{equation}
The contour orientation supplies the sign of the coherent term.  
The branch factor $s(z)$ gives the opposite anti-Hermitian signs on forward and backward evolution.  
The jump vertex contains operators at the same real time on opposite branches, and $\theta_-(z)$ includes it once to avoid double counting.

The equivalence between Eq.~\eqref{eq:SM_KL_contour_generator} and the Lindblad evolution can be shown compactly by expanding in the jump term.  
Introduce the non-Hermitian propagators
\begin{equation}
U(t,t')
=
T\exp\!\left[
-i\int_{t'}^{t}dt_1H_{\rm NH}(t_1)
\right],
\qquad
U^\dagger(t,t')
=
\bar T\exp\!\left[
i\int_{t'}^{t}dt_1H_{\rm NH}^\dagger(t_1)
\right],
\label{eq:SM_KL_nh_propagators}
\end{equation}
where $T$ and $\bar T$ denote real-time ordering and anti-ordering. 
The corresponding quantum-jump expansion is
\begin{equation}
\rho(t)
=
\sum_{n=0}^{\infty}
\sum_{\mu_1\dots\mu_n}
\int\limits_{t>t_1>\dots>t_n>t_0}\!\!
dt_1\cdots dt_n\,
\mathcal{J}_n\,
\rho(t_0)\,
\mathcal{J}_n^\dagger,
\label{eq:SM_KL_jump_expansion}
\end{equation}
where
\begin{equation*}
\mathcal{J}_n
\equiv
\mathcal{J}_n(\mu_1t_1,\dots,\mu_nt_n)
=
U(t,t_1)L_{\mu_1}U(t_1,t_2)
\cdots
L_{\mu_n}U(t_n,t_0).
\end{equation*}
Here $n$ is the number of jumps, the integral orders their times, and $\mathcal{J}_0=U(t,t_0)$.  
Contour ordering reproduces this expansion without placing the operators on the two sides of $\rho$ by hand.  
Indeed, writing $t_1^{\mp}$ for real time $t_1$ on $C_{\mp}$, the contour orientation gives
\begin{equation}
-i\int_C dz_1H_C(z_1,z_1^*)
=
-i\int_{t_0}^{t}\!dt_1\,H_C(t_1^-,t_1^+)
+i\int_{t_0}^{t}\!dt_1\,H_C(t_1^+,t_1^-).
\label{eq:SM_KL_contour_split}
\end{equation}
Using $s(t_1^\mp)=\pm1$, $\theta_-(t_1^-)=1$, and $\theta_-(t_1^+)=0$, the two branch generators are
\begin{subequations}
\begin{align}
H_C(t_1^-,t_1^+)
&=
H_{\rm NH}(t_1)
+
i\sum_\mu L_\mu^\dagger(t_1^+)L_\mu(t_1^-),
\label{eq:SM_KL_generator_forward}
\\
H_C(t_1^+,t_1^-)
&=
H_{\rm NH}^\dagger(t_1),
\label{eq:SM_KL_generator_backward}
\end{align}
\end{subequations}
with $H_{\rm NH}$ defined in Eq.~\eqref{eq:SM_KL_nonhermitian_H}.  
Substitution into Eq.~\eqref{eq:SM_KL_contour_split} yields
\begin{equation}
-i\int_{t_0}^{t}\!dt_1\,H_{\rm NH}(t_1)
\;+\;
i\int_{t_0}^{t}\!dt_1\,H_{\rm NH}^\dagger(t_1)
\;+\;
\int_{t_0}^{t}\!dt_1\sum_\mu L_\mu^\dagger(t_1^+)L_\mu(t_1^-).
\label{eq:SM_KL_contour_exponent}
\end{equation}
The first two terms generate $U$ on the ket and $U^\dagger$ on the bra.  
Expanding the third term inserts cross-branch jump vertices.  
Each vertex is even under fermion parity, so the $n!$ equivalent time orderings cancel the factor $1/n!$ and give the ordered product $\mathcal{J}_n\rho(t_0)\mathcal{J}_n^\dagger$ in Eq.~\eqref{eq:SM_KL_jump_expansion}.  
Thus the contour exponential reproduces the Lindblad evolution exactly.

We thus obtain the contour representation of an expectation value,
\begin{equation}
\langle O(z)\rangle
=
{\rm Tr}\!
\left[
\rho(t_0)
T_C\!\left[
\exp\!\left(
-i\int_C dz_1H_C(z_1,z_1^*)
\right)
O(z)
\right]
\right].
\label{eq:SM_KL_contour_expectation}
\end{equation}
Equation~\eqref{eq:SM_KL_contour_expectation} retains both the branch-diagonal no-jump evolution and the branch-mixing jump processes, and is the starting point for the Green's-function construction below. 
For the contour Green's function
\begin{equation}
G_{ij}(z,z')
=
-i
\left\langle
T_C d_i(z)d_j^\dagger(z')
\right\rangle,
\label{eq:SM_KL_contour_green}
\end{equation}
differentiation with respect to $z$ gives
\begin{equation}
i\partial_zG_{ij}(z,z')
=
\delta_{ij}\delta_C(z,z')
+
D^{(H)}_{ij}
+
D^{({\rm nj})}_{ij}
+
D^{({\rm jump})}_{ij},
\label{eq:SM_KL_contour_eom}
\end{equation}
where
\begin{subequations}\label{eq:SM_KL_eom}
\begin{align}
D^{(H)}_{ij}
&=
-i
\left\langle
T_C [d_i(z),H(z)]d_j^\dagger(z')
\right\rangle,
\label{eq:SM_KL_coherent_eom}\\
D^{({\rm nj})}_{ij}
&=
-\frac{1}{2}s(z)
\sum_\mu
\left\langle
T_C [d_i(z),L_\mu^\dagger(z)L_\mu(z)]d_j^\dagger(z')
\right\rangle,
\label{eq:SM_KL_nojump_eom}\\
D^{({\rm jump})}_{ij}
&=
-\theta_-(z)
\sum_\mu
\left\langle
T_C
L_\mu^\dagger(z^*)\{d_i(z),L_\mu(z)\}
d_j^\dagger(z')
\right\rangle
-
\theta_+(z)
\sum_\mu
\left\langle
T_C
\{d_i(z),L_\mu^\dagger(z)\}L_\mu(z^*)
d_j^\dagger(z')
\right\rangle.
\label{eq:SM_KL_jump_eom}
\end{align}
\end{subequations}
Here $\delta_C(z,z')$ is the delta function on the contour, $D^{(H)}$ is the coherent contribution, $D^{({\rm nj})}$ is the damping from the anti-Hermitian part of $H_{\rm NH}$, and $D^{({\rm jump})}$ is the jump contribution. 
The anticommutators in $D^{({\rm jump})}$ follow from moving a fermionic Lindblad operator past $d_i$.

\subsection{Quadratic Hamiltonian and linear Lindblad operators}
\label{sec:SM_KL_quadratic}

We now evaluate the contour self-energies for the case solved in this work: a quadratic Hamiltonian with linear Lindblad channels. 
Let
\begin{equation}
H
=
\sum_{ij}d_i^\dagger h_{ij}d_j,
\label{eq:SM_KL_quadratic_hamiltonian}
\end{equation}
and include linear loss and gain Lindblad operators
\begin{equation}
L_{\ell,\mu}
=
\sum_i a_{\mu i}d_i,
\qquad
L_{g,\nu}
=
\sum_i b_{\nu i}d_i^\dagger.
\label{eq:SM_KL_linear_channels}
\end{equation}
The operators $L_{\ell,\mu}$ and $L_{g,\nu}$ remove and add a particle, respectively.  
The engineered reservoir has no physical particle-gain process, but we keep both algebraic cases: their contrast makes the distribution components explicit, and after the particle-hole transformation to Nambu space the same physical spin-down electron loss appears as hole gain (Sec.~\ref{sec:SM_KL_nambu}). 
The canonical anticommutation relations give
\begin{equation}
[d_i,H]=\sum_k h_{ik}d_k,
\qquad
[d_i,L_{\ell,\mu}^\dagger L_{\ell,\mu}]=\sum_k a_{\mu i}^*a_{\mu k}d_k,
\qquad
[d_i,L_{g,\nu}^\dagger L_{g,\nu}]=-\sum_k b_{\nu i}b_{\nu k}^*d_k.
\label{eq:SM_KL_commutators}
\end{equation}
Equation~\eqref{eq:SM_KL_commutators} shows explicitly that the operator generated by each commutator is linear in the fermion fields.  
The anticommutators in $D^{({\rm jump})}$ are c-numbers and likewise leave one linear Lindblad operator. 
After multiplication by the external field $d_j^\dagger$, every term in Eq.~\eqref{eq:SM_KL_contour_eom} is a two-point correlator. 
The equations of motion thus close at the one-particle level without requiring the Zeno or infinite-gap limit.
Substituting Eqs.~\eqref{eq:SM_KL_linear_channels} and \eqref{eq:SM_KL_commutators} into the contour equation of motion, Eq.~\eqref{eq:SM_KL_contour_eom}, and resolving the result into its real-time components gives the Lindblad self-energies
\begin{equation}
\Sigma^R_{\rm Lind}
=
-\frac{i}{2}\left(\ell^>+\ell^<\right),
\qquad
\Sigma^A_{\rm Lind}
=
+\frac{i}{2}\left(\ell^>+\ell^<\right),
\qquad
\Sigma^<_{\rm Lind}
=
i\ell^<,
\qquad
\Sigma^>_{\rm Lind}
=
-i\ell^>,
\label{eq:SM_KL_lindblad_selfenergies}
\end{equation}
where we have defined the Hermitian loss and gain rate matrices
\begin{equation}
\ell^>_{ij}
=
\sum_\mu a_{\mu i}^*a_{\mu j},
\qquad
\ell^<_{ij}
=
\sum_\nu b_{\nu i}b_{\nu j}^*.
\label{eq:SM_KL_rate_matrices}
\end{equation}
The self-energies obey $\Sigma^>_{\rm Lind}-\Sigma^<_{\rm Lind}=\Sigma^R_{\rm Lind}-\Sigma^A_{\rm Lind}$ and they are frequency independent since the Lindblad reservoir is Markovian. 
For a pure-loss reservoir in an ordinary particle basis, $\ell^<=0$.  
We denote this specialization of Eq.~\eqref{eq:SM_KL_lindblad_selfenergies} by $\Sigma^X_{\rm loss}\equiv\left.\Sigma^X_{\rm Lind}\right|_{\ell^<=0}$, giving
\begin{equation}
\Sigma^R_{\rm loss}
=
-\frac{i}{2}\ell^>,
\qquad
\Sigma^A_{\rm loss}
=
+\frac{i}{2}\ell^>,
\qquad
\Sigma^<_{\rm loss}=0,
\qquad
\Sigma^>_{\rm loss}
=
-i\ell^>.
\label{eq:SM_KL_pure_loss_selfenergy}
\end{equation}

Given the Lindblad self-energies in Eq.~\eqref{eq:SM_KL_lindblad_selfenergies}, the steady-state retarded and advanced Green's functions follow from the Dyson equation:
\begin{equation}
G^R(\omega)
=
\left[
\omega+i0^+-h-\Sigma^R_{\rm Lind}(\omega)
\right]^{-1},
\qquad
G^A(\omega)=\left[G^R(\omega)\right]^\dagger,
\label{eq:SM_KL_quadratic_retarded_green}
\end{equation}
and the lesser and greater components follow from the Keldysh formula
\begin{equation}
G^<(\omega)
=
G^R(\omega)\Sigma^<_{\rm Lind}(\omega)G^A(\omega),
\qquad
G^>(\omega)
=
G^R(\omega)\Sigma^>_{\rm Lind}(\omega)G^A(\omega).
\label{eq:SM_KL_quadratic_keldysh_green}
\end{equation}
Equation~\eqref{eq:SM_KL_quadratic_keldysh_green} is the stationary Keldysh solution after the initial-state term has decayed.  
This requires every mode to acquire a finite linewidth and the steady state to be unique.

\subsection{Finite-gap steady state of the double-dot junction}
\label{sec:SM_KL_nambu}

We now specialize the quadratic Keldysh construction to the finite-gap double-dot junction. 
We define a four-component Nambu spinor
\begin{equation}
\Phi
=
\left(
d_{L\uparrow},
d_{L\downarrow}^\dagger,
d_{R\uparrow},
d_{R\downarrow}^\dagger
\right)^T.
\label{eq:SM_KL_nambu_spinor}
\end{equation}
In the finite-gap formulation, we retain the hopping Hamiltonian $H_t$ of Eq.~\eqref{eq:SM_Ht} but allow general on-site energies through $H_\epsilon=\sum_{j\sigma}\epsilon_j d_{j\sigma}^\dagger d_{j\sigma}$. 
Superconducting leads are represented by the frequency-dependent self-energies below, rather than by the static pairing Hamiltonian $H_\Delta$ of Eq.~\eqref{eq:SM_HDelta}. 
Thus the bare-dot Hamiltonian $H_{\rm dot}=H_t+H_\epsilon$ becomes
\begin{equation}
H_{\rm dot}
=
\Phi^\dagger h_{\rm dot} \Phi
+
{\rm const.}~,
\label{eq:SM_KL_dot_hamiltonian_nambu}
\end{equation}
where
\begin{equation}
h_{\rm dot}
=
\begin{pmatrix}
\epsilon_L & 0 & t & 0\\
0 & -\epsilon_L & 0 & -t\\
t & 0 & \epsilon_R & 0\\
0 & -t & 0 & -\epsilon_R
\end{pmatrix}.
\label{eq:SM_KL_dot_matrix}
\end{equation}
Here $\epsilon_j$ is the on-site energy of dot $j=L,R$, and $t$ is real in the gauge of Sec.~\ref{sec:SM_bright_dark}.

The BCS retarded/advanced lead self-energy in the wide-band limit reads
\begin{equation}
\Sigma^R_\alpha(\omega)
=
-\Gamma_\alpha
\frac{
\omega\tau_0
+
\Delta_\alpha
\left(
\cos\phi_\alpha\tau_x
-
\sin\phi_\alpha\tau_y
\right)
}{
\sqrt{\Delta_\alpha^2-(\omega+i0^+)^2}
},
\qquad
\Sigma^A_\alpha(\omega)
=
\left[\Sigma^R_\alpha(\omega)\right]^\dagger.
\label{eq:SM_KL_lead_retarded/advanced_selfenergy}
\end{equation}
Here $\tau_0$ is the identity and $\tau_{x,y,z}$ are Pauli matrices in Nambu space.
Since the leads are assumed at local equilibrium, the corresponding lesser and greater components obey the fluctuation-dissipation theorem:
\begin{subequations}
\begin{align}
\Sigma^<_\alpha(\omega)
&=
f(\omega)
\left[
\Sigma^A_\alpha(\omega)-\Sigma^R_\alpha(\omega)
\right],
\label{eq:SM_KL_lead_lesser_selfenergy}\\
\Sigma^>_\alpha(\omega)
&=
\left[f(\omega)-1\right]
\left[
\Sigma^A_\alpha(\omega)-\Sigma^R_\alpha(\omega)
\right],
\label{eq:SM_KL_lead_greater_selfenergy}
\end{align}
\end{subequations}
where $f(\omega)$ is the common equilibrium Fermi function of the leads.

The engineered loss enters through its own self-energy. 
The nonlocal jump operator in Eq.~\eqref{eq:SM_nonlocal_jump} describes pure electron loss, so Eq.~\eqref{eq:SM_KL_pure_loss_selfenergy} applies in an ordinary particle basis. 
In the site order $(L,R)$, writing $L_\uparrow=\sum_i a_{\uparrow i}d_{i\uparrow}$ gives $a_\uparrow=\sqrt{\kappa/2}(e^{i\chi},1)$. 
The loss definition $\ell^>_{ij}=a_i^*a_j$ in Eq.~\eqref{eq:SM_KL_rate_matrices} then yields
\begin{equation}
\ell^>_\uparrow
=
\frac{\kappa}{2}
\begin{pmatrix}
1&e^{-i\chi}\\
e^{i\chi}&1
\end{pmatrix}.
\label{eq:SM_KL_electron_loss_rate}
\end{equation}
For the spin-down block, the Nambu fields are holes, $h_{i\downarrow}=d_{i\downarrow}^\dagger$, and the same physical jump becomes $L_\downarrow=\sum_i b_{\downarrow i}h_{i\downarrow}^\dagger$ with $b_\downarrow=a_\uparrow$.  
The gain definition $\ell^<_{ij}=b_i b_j^*$ in Eq.~\eqref{eq:SM_KL_rate_matrices} reverses the order of complex conjugation and gives
\begin{equation}
\ell^<_\downarrow
=
\left(\ell^>_\uparrow\right)^*
=
\frac{\kappa}{2}
\begin{pmatrix}
1&e^{i\chi}\\
e^{-i\chi}&1
\end{pmatrix}.
\label{eq:SM_KL_hole_gain_rate}
\end{equation}
Embedding these blocks in the four-component order of Eq.~\eqref{eq:SM_KL_nambu_spinor} and substituting them into the general Lindblad result, Eq.~\eqref{eq:SM_KL_lindblad_selfenergies}, gives the retarded/advanced components
\begin{equation}
\Sigma^R_{\rm loss}
=
-\frac{i\kappa}{4}
\begin{pmatrix}
1&0&e^{-i\chi}&0\\
0&1&0&e^{i\chi}\\
e^{i\chi}&0&1&0\\
0&e^{-i\chi}&0&1
\end{pmatrix},
\qquad
\Sigma^A_{\rm loss}
=
\left(\Sigma^R_{\rm loss}\right)^\dagger
=
-\Sigma^R_{\rm loss},
\label{eq:SM_KL_loss_spectral_nambu}
\end{equation}
and lesser/greater components
\begin{equation}
\Sigma^<_{\rm loss}
=
i\frac{\kappa}{2}
\begin{pmatrix}
0&0&0&0\\
0&1&0&e^{i\chi}\\
0&0&0&0\\
0&e^{-i\chi}&0&1
\end{pmatrix},
\qquad
\Sigma^>_{\rm loss}
=
-i\frac{\kappa}{2}
\begin{pmatrix}
1&0&e^{-i\chi}&0\\
0&0&0&0\\
e^{i\chi}&0&1&0\\
0&0&0&0
\end{pmatrix}.
\label{eq:SM_KL_loss_distribution_nambu}
\end{equation}
The identity $\Sigma^>_{\rm loss}-\Sigma^<_{\rm loss}=\Sigma^R_{\rm loss}-\Sigma^A_{\rm loss}$ holds exactly.

We now assemble the finite-gap steady state. 
Since the leads and the reservoir are independent, the total self-energy reads:
\begin{equation}
\Sigma^X_{\rm tot}(\omega)
=
\Sigma^X_L(\omega)
+
\Sigma^X_R(\omega)
+
\Sigma^X_{\rm loss},
\qquad
X=R,A,<,>.
\label{eq:SM_KL_total_selfenergy}
\end{equation}
The retarded Green's function is obtained by the Dyson equation
\begin{equation}
G^R(\omega)
=
\left[
(\omega+i0^+)\mathbb{I}
-
h_{\rm dot}
-
\Sigma^R_{\rm tot}(\omega)
\right]^{-1}.
\label{eq:SM_KL_dyson_retarded}
\end{equation}
Its poles give the Andreev resonances of the open junction.  
At large $\kappa$, $\Sigma^R_{\rm loss}$ separates a broad bright branch from a narrow dark branch, recovering the scale separation underlying the Zeno elimination in Sec.~\ref{sec:SM_zeno_elimination}.  
The crossover is continuous at finite $\kappa$. 
The advanced component is $ G^A(\omega) =
\left[G^R(\omega)\right]^\dagger
$. 
The stationary distribution then follows from the total lesser and greater self-energies of Eq.~\eqref{eq:SM_KL_total_selfenergy},
\begin{equation}
G^\lessgtr(\omega)
=
G^R(\omega)\Sigma^\lessgtr_{\rm tot}(\omega)G^A(\omega).
\label{eq:SM_KL_dyson_lesser_greater}
\end{equation}

Equations~\eqref{eq:SM_KL_dyson_retarded} and \eqref{eq:SM_KL_dyson_lesser_greater} define the finite-gap steady state used in Fig.~\ref{fig:robustness} of the main text.

\subsection{Lead current, loss current, and continuity relation}
\label{sec:SM_KL_currents}

We use $e=1$ in this section, so charge currents are expressed as electron-number currents. The lead current $I_\alpha$, taken positive when electrons flow from lead $\alpha=L,R$ into the double dot, follows from the time derivative of the lead particle number before that lead is integrated out.
In the stationary state it is
\begin{equation}
I_\alpha
=
\int\frac{d\omega}{2\pi}
{\rm Tr}
\left\{
Q
\left[
G^R(\omega)\Sigma_\alpha^<(\omega)
+
G^<(\omega)\Sigma_\alpha^A(\omega)
-
\Sigma_\alpha^R(\omega)G^<(\omega)
-
\Sigma_\alpha^<(\omega)G^A(\omega)
\right]
\right\},
\label{eq:SM_KL_lead_current_keldysh}
\end{equation}
where $Q={\rm diag}(1,-1,1,-1)$ is the particle-number metric.
Using the Hermiticity relations of the Green's functions and self-energies gives the equivalent compact form
\begin{equation}
I_\alpha
=
2{\rm Re}
\int\frac{d\omega}{2\pi}
{\rm Tr}
\left\{
Q
\left[
G^R(\omega)\Sigma_\alpha^<(\omega)
+
G^<(\omega)\Sigma_\alpha^A(\omega)
\right]
\right\}.
\label{eq:SM_KL_lead_current_realpart}
\end{equation}
Equations~\eqref{eq:SM_KL_lead_current_keldysh} and \eqref{eq:SM_KL_lead_current_realpart} retain the coherent condensate contribution carried by the anomalous lead self-energy. 
In particular, in the infinite-gap limit $\Sigma_\alpha^{</>}\to0$ and $\Sigma_\alpha^R=\Sigma_\alpha^A$ at finite frequency, but the terms $G^<\Sigma_\alpha^A-\Sigma_\alpha^R G^<$ remain finite and reduce to $I_\alpha=-2\langle\partial_{\phi_\alpha}H_\Delta\rangle$. 
A normal-state collision form with $Q[\Sigma_\alpha^<G^>-\Sigma_\alpha^>G^<]$ cannot be used here because the anomalous superconducting self-energy does not commute with $Q$.

Charge can also leave through the engineered reservoir.  
The total loss current is the number of particles removed per Lindblad event times the event rate,
\begin{equation}
I_{\rm loss,tot}
=
\sum_\sigma
\left\langle
L_\sigma^\dagger L_\sigma
\right\rangle.
\label{eq:SM_KL_total_loss_current}
\end{equation}
With $d_e=(d_{L\uparrow},d_{R\uparrow},d_{L\downarrow},d_{R\downarrow})^T$,
\begin{equation}
\sum_\sigma L_\sigma^\dagger L_\sigma
=
d_e^\dagger \ell_e^>d_e,
\qquad
\ell_e^>
=
\frac{\kappa}{2}
\begin{pmatrix}
1&e^{-i\chi}&0&0\\
e^{i\chi}&1&0&0\\
0&0&1&e^{-i\chi}\\
0&0&e^{i\chi}&1
\end{pmatrix}.
\label{eq:SM_KL_loss_matrix_electron}
\end{equation}
Therefore
\begin{equation}
I_{\rm loss,tot}
=
{\rm Tr}_e\left[\ell_e^>\rho_e\right].
\label{eq:SM_KL_total_loss_green}
\end{equation}
The electron density matrix is obtained from the equal-time Nambu density matrix $\rho_N=-i\int (d\omega/2\pi)\,G^<(\omega)$.  
In the site order $(L,R)$, let $\rho_\uparrow=(\rho_N)_{\{1,3\},\{1,3\}}$ and $\rho_h=(\rho_N)_{\{2,4\},\{2,4\}}$.  
The hole block satisfies $(\rho_h)_{ij}=\delta_{ij}-(\rho_\downarrow)_{ji}$, so $\rho_e=\rho_\uparrow\oplus\left(\mathbb I_2-\rho_h^T\right)$. 
Equations~\eqref{eq:SM_KL_loss_matrix_electron} and \eqref{eq:SM_KL_total_loss_green} use only the same $4\times4$ Nambu Green's function as the lead-current calculation.

The continuity equation is most directly obtained before the leads are integrated out.  
For $\hat{N}_C=\sum_{i=L,R}\sum_\sigma d^\dagger_{i\sigma}d_{i\sigma}$, the tunneling Hamiltonian gives the two lead currents and the Lindblad term removes particles at the rate in Eq.~\eqref{eq:SM_KL_total_loss_current}.  
Therefore
\begin{equation}
\frac{d}{dt}\langle \hat{N}_C\rangle
=
I_L+I_R-I_{\rm loss,tot}.
\label{eq:SM_KL_continuity_time}
\end{equation}
In the steady state the left-hand side vanishes and
\begin{equation}
I_L+I_R=I_{\rm loss,tot}.
\label{eq:SM_KL_continuity_steady}
\end{equation}
Thus $I_L$ and $I_R$ need not be equal and opposite.  
Their sum supplies the loss channel.  
We decompose the terminal currents into transport and common-mode components,
\begin{equation}
I_{\rm tr}
=
\frac{I_L-I_R}{2},
\qquad
I_{\rm cm}
=
\frac{I_L+I_R}{2}.
\label{eq:SM_KL_current_decomposition}
\end{equation}
Equivalently, $I_L=I_{\rm cm}+I_{\rm tr}$ and $I_R=I_{\rm cm}-I_{\rm tr}$.  
The transport component contributes with opposite signs at the two terminals, whereas the common-mode component contributes equally to both.  
The factors $1/2$ normalize this decomposition so that a lossless current $I_L=-I_R=J$ gives $I_{\rm tr}=J$. They do not originate from the Cooper-pair charge $2e$.  
The transport current $I_{\rm tr}$ is conjugate to the superconducting phase difference and is used to characterize the diode response.  
In the lossless case, $I_{\rm tr}>0$ corresponds to left-to-right transport.  
In a general open three-terminal junction, unequal left and right leakage can also contribute to $I_{\rm tr}$, so it does not by itself isolate a lead-to-lead supercurrent.

For direct comparison with the common-mode current, we define the normalized half-total outflow
\begin{equation}
I_{\rm loss}
\equiv
\frac{I_{\rm loss,tot}}{2},
\qquad
I_{\rm cm}=I_{\rm loss}.
\label{eq:SM_KL_normalized_loss}
\end{equation}
The quantity $I_{\rm loss}$ is one half of the physical total reservoir outflow, and in the spin-symmetric problem it is also the per-spin outflow.
The equality $I_{\rm cm}=I_{\rm loss}$ is Eq.~\eqref{eq:SM_KL_continuity_steady} divided by two.
In the leading-order noninteracting infinite-gap Zeno theory, $I_{\rm cm}=O(\kappa^{-1})$ is neglected and $I_{\rm tr}$ reduces to the secular current $I_{\rm ss}$ in Eq.~\eqref{eq:SM_factorized_current}.

The two forms in Eqs.~\eqref{eq:SM_KL_lead_current_keldysh} and \eqref{eq:SM_KL_lead_current_realpart} must agree, the infinite-gap limit must reproduce the phase-derivative current, and the resulting currents must satisfy Eq.~\eqref{eq:SM_KL_continuity_steady}.  
Together these checks probe both the spectral and distribution components of the numerical solution.

\subsection{Finite-$\kappa$ diode efficiency}
\label{sec:SM_operating_window}

To isolate the finite-$\kappa$ crossover from the finite-gap corrections of Secs.~\ref{sec:SM_KL_nambu}-\ref{sec:SM_KL_currents}, we return to the infinite-gap model.  
At $\Delta_\alpha=\infty$ the lead self-energies reduce to frequency-independent coherent pairing terms and their quasiparticle lesser and greater components vanish at finite frequency.  
The steady state therefore follows directly from the Lindblad equation in the full $2^4=16$-dimensional Fock space, without eliminating the bright mode and without frequency integrals.  
This result retains the full bright and dark Fock space and is exact in $\kappa$, providing a benchmark for the projected dark-sector theory of Sec.~\ref{sec:SM_projected_theory} without adjustable parameters.
The lead-resolved currents are evaluated from $I_\alpha=-2\langle\partial_{\phi_\alpha}H_\Delta\rangle$ with the sign convention of Sec.~\ref{sec:SM_KL_currents}.  
For each $(\chi,\kappa)$ we then form $I(\phi)\equiv I_{\rm tr}(\phi)$ and extract the critical currents $I_c^\pm$ and the diode efficiency $\eta_D$ from Eqs.~\eqref{eq:SM_Ic_asymmetry} and \eqref{eq:SM_etaD_asymmetry}, now with $I_{\rm tr}$ in place of $I_{\rm ss}$.  
Thus $I_c^+=0$ for an entirely negative CPR, while $|I_c^-|=0$ for an entirely positive one.

\begin{figure}
\centering
\includegraphics[width=\linewidth]{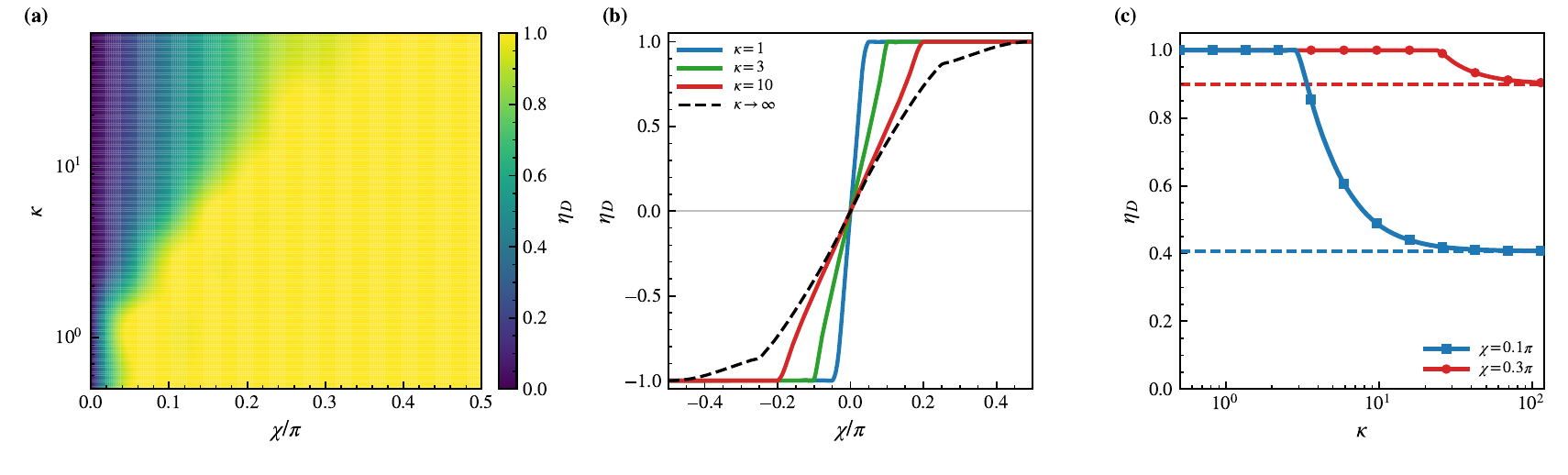}
\caption{
Finite-$\kappa$ diode efficiency from the direct Lindblad solution at $t=\Gamma_L=\Gamma_R=1$ and zero on-site energies, $\epsilon_L=\epsilon_R=0$.
(a) Diode efficiency $\eta_D$ in the $(\chi,\kappa)$ plane.  
The broad region with $\eta_D\simeq1$ shows that the backward-current branch is strongly suppressed relative to the forward branch. 
(b) $\eta_D$ versus reservoir phase $\chi$ for $\kappa=1,3,10$, together with the analytic Zeno result of Eq.~\eqref{eq:SM_Iss_closed} ($\kappa\to\infty$, dashed).  
Reversing $\chi$ reverses the diode polarity, $\eta_D(-\chi)=-\eta_D(\chi)$. 
(c) $\eta_D$ versus $\kappa$ at fixed $\chi=0.1\pi$ and $0.3\pi$.  
Dashed lines denote the corresponding $\kappa\to\infty$ values.
}
\label{fig:operating_window}
\end{figure}

Figure~\ref{fig:operating_window} shows a broad finite-$\kappa$ regime with strong directional asymmetry.  
In part of this region the CPR has one sign over the full phase cycle and $|\eta_D|=1$.  
The exact result approaches the analytic Zeno-limit curve continuously as $\kappa$ increases.  
Exchanging the left and right dots maps $(\phi,\chi)$ to $(-\phi,-\chi)$ and reverses the current, so $I(\phi,\chi)=-I(-\phi,-\chi)$ and $\eta_D(-\chi)=-\eta_D(\chi)$.

The Zeno effective theory is the strong-dissipation limit of this finite-$\kappa$ response.  
The finite-$\kappa$ calculations connect continuously to the occupation-reweighting mechanism derived in Sec.~\ref{sec:SM_current}. The occupation-only attribution and factorized current apply to the leading noninteracting infinite-gap Zeno theory under the secular condition stated there.

\end{document}